\documentclass[structabstract]{aa}   
\usepackage{graphicx}
\usepackage{times}
\usepackage{epsfig}
\usepackage{lscape}
\usepackage{amssymb}
\usepackage{latexsym}
\usepackage{natbib}
\usepackage{txfonts}
\def\kms{$\rm km\;s^{-1}$}

\def\ha{H$\alpha$}
\def\hb{H$\beta$}
\def\hg{H$\gamma$}
\def\nipg{[N~{\small I}]$\,\lambda\lambda5198,5200$} 
\def\oiiipg{[O~{\small III}]$\,\lambda\lambda4959,5007$}
\def\oiiip{[O~{\small III}]$\,\lambda4959$}

\def\oiii{[\ion{O}{iii}]}
\def\nii{[N~{\small II}]}

\def\sii{[S~{\small II}]}

\def\mage{$\overline{{\rm Age}}$}
\def\mfe{$\overline{{\rm [Z/H]}}$}
\def\malpha{$\overline{{\rm [}\alpha{\rm /Fe}]}$}
\def\mgb{Mg~{\it b}}
\def\fei{Fe{\small 5270}}
\def\feii{Fe{\small 5335}}

\begin{document}
\title{Spectroscopic evidence of distinct stellar populations in the counter-rotating
stellar disks of NGC 3593 and NGC 4550.
\thanks{Based on observations
    collected at the European Southern Observatory for the programme
    087.B-0853A.}}
   \subtitle{ }

   \author{L. Coccato \inst{1}
     \and L. Morelli \inst{2,3} 
     \and A. Pizzella \inst{2,3} 
     \and E. M. Corsini \inst{2,3}
     \and L. M. Buson \inst{3} 
     \and E. Dalla Bont\`a \inst{2,3} 
          }
  \offprints{L. Coccato, e-mail: lcoccato@eso.org}

  \institute{European Southern Observatory,
    Karl-Schwarzschild-Stra$\beta$e 2, D-85748 Garching bei M\"unchen,
    Germany.  
    \and Dipartimento di Fisica e Astronomia ``G. Galilei'',
    Universit\`a di Padova, vicolo dell'Osservatorio 3, I-35122
    Padova, Italy. 
    \and INAF Osservatorio Astronomico di Padova, vicolo dell'Osservatorio 5, I-35122 Padova, Italy.}

   \date{Received ...; accepted ...}


\abstract
   {}
  {We present the results of integral-field spectroscopic observations
    of the two disk galaxies NGC~3593 and NGC~4550 obtained with the
    Visible Multi Object Spectrograph at the Very Large
    Telescope. Both galaxies are known to host two counter-rotating
    stellar disks, with the ionized gas co-rotating with one of
    them. We measured in each galaxy the surface brightness,
    kinematics, mass surface density, and the stellar populations of
    the two stellar components as well as the distribution,
    kinematics, and metallicity of the ionized-gas component to
    constrain the formation scenario of these peculiar galaxies.}
  {We applied a novel spectroscopic decomposition technique to both
  galaxies, to disentangle at each position in the field of view the
  relative contribution of the two counter-rotating stellar and one
  ionized-gas components to the observed spectrum. We measured the
  kinematics and the line strengths of the Lick indices of the two
  counter-rotating stellar components. We modeled the data of each
  stellar component with single stellar population models that account
  for the $\alpha$/Fe overabundance.}
{In both galaxies we successfully separated the main from the
  secondary stellar component that is less massive and rotates in the
  same direction of the ionized-gas component. The two stellar
  components have exponential surface-brightness profiles. In NGC~3593
  they have different scale lengths, with the secondary one dominating
  the innermost 500 pc. In NGC~4550 they have the same scale lengths,
  but slightly different scale heights. In both galaxies, the two
  counter-rotating stellar components have different stellar
  populations. The secondary stellar disk is younger, more metal poor,
  and more $\alpha$-enhanced than the main galaxy stellar disk. Such a
  difference is stronger in NGC~3593 than in NGC~4550.}
{Our findings rule out an internal origin of the secondary stellar
  component and favor a scenario where it formed from gas accreted on
  retrograde orbits from the environment fueling an in situ outside-in
  rapid star formation. The event occurred $\simeq\,2$ Gyr ago in NGC
  3593 ($1.6 \pm 0.8$ Gyr after the formation of the main galaxy
  disk), and $\simeq 7$ Gyr ago in NGC 4550, (less than 1 Gyr after the
  formation of the main galaxy disk). The formation through a binary
  galaxy merger cannot be ruled out, and a larger sample is required
  to statistically determine which is the most efficient mechanism to
  build counter-rotating stellar disks.}


  \titlerunning{Distinct stellar populations in the counter-rotating
    stellar disks of NGC 3593 and NGC 4550.}

  \authorrunning{L. Coccato et al.}

   \maketitle
%

\section{Introduction}

The general term of ``counter-rotation'' indicates the presence of two
components of a galaxy that rotate along opposite directions with
respect to each other.  This phenomenon has been observed in galaxies
all over the Hubble sequence, from ellipticals to irregulars,
including barred galaxies.
Counter-rotating galaxies are classed depending on the nature (stars
vs. stars, stars vs. gas, gas vs. gas) and size
(counter-rotating cores, rings, disks) of the counter-rotating
components (see \citealt{Rubin94, Galletta96, Bertola+99b} for
reviews).

In this paper, we will focus on the class of disk galaxies with two
counter-rotating stellar disks of comparable size. To date, the few
known objects are: NGC~4550 \citep{Rubin+92, Rix+92, Emsellem+04},
NGC~7217 \citep{Merrifield+94}, NGC~3593 \citep{Bertola+96,
  Corsini+98, Corsini+99, Garcia+00}, NGC~4138 \citep{Jore+96, Haynes+00}, and
NGC~5719 \citep{Vergani+07, Coccato+11}.

The current paradigm that explains stellar counter-rotation is a
retrograde acquisition of external gas and subsequent star formation
\citep{Pizzella+04}. This scenario explains also the lower detection
rate of counter-rotation in spirals with respect to S0's: a
counter-rotating gaseous disk will be observed only if the mass of the
newly supplied gas exceeds that of the pre-existing
one \citep{Lovelace+96, Thakar+96}, which is higher in spirals than in
S0's.
About 30 per cent of S0's host a counter-rotating gaseous disk
\citep{Bertola+92, Kuijken+96, Bureau+06}, although that less than
 10 per cent of them have a significant fraction of counter-rotating stars
\citep{Kuijken+96}. For comparison, less than 10 per cent of spiral
galaxies are found to host a counter-rotating gaseous and/or stellar
disk \citep{Kannappan+01, Pizzella+04}.

Alternative scenarios for building counter-rotating stellar disks are
also proposed. They include the accretion of already formed stars (and
gas) through mergers and internal secular processes related to disk
instabilities.
Indeed, galaxy mergers can produce galaxies with counter-rotating
components, depending on the orbital parameters of the encounter and
the amount of gas involved in the process. The merger remnants range
from kinematically decoupled cores or counter-rotating bulges in the
case of minor mergers (e.g., \citealt{Balcells+98, Jesseit+07,
  Eliche-Moral+11, Bois+11}), to large-scale stellar disks in the case
of binary major mergers (e.g., \citealt{Puerari+01, Crocker+09}).
An alternative to the external-origin scenarios has been proposed by
\citet{Evans+94} and it involves the dissolution of a triaxial
potential or a bar. In this process, the stars moving on box orbits
escape from the confining azimuthal potential well to move onto tube
orbits. In non-rotating disks, there are as many box orbits with
clockwise azimuthal motion as counter-clockwise. Thus, half box-orbit
stars are scattered onto clockwise streaming tube orbits, half onto
counter-clockwise. In this way, two identical counter-rotating stellar
disks can be built.
Barred galaxies host quasi-circular retrograde orbits
\citep[see][]{Wozniak+97}. Although the origin of stellar
counter-rotation observed in barred galaxies \citep{Bettoni+89,
  Bettoni+97} is not necessarily external,  acquired gas can be
trapped on this family of retrograde orbits and then eventually form
stars.

The different formation mechanisms are expected to leave different
signatures in the properties of the stellar populations of the
counter-rotating component. In particular, the age is a key element to
investigate the origin of counter-rotating stars.
Gas acquisition followed by star formation predicts younger ages for
the counter-rotating stellar component in {\it all\/} cases, and
it allows for different metallicity and $\alpha$-enhancement between
the two disks, which are formed by different gas.
Direct acquisition of stars through mergers also allows for different
metallicity and $\alpha$-enhancement, but the younger component will
be determined by the difference in age between the host galaxy and
merged system. Roughly, one would expect younger counter-rotating
stars in $\sim50$ per cent of the cases.
On the contrary, the internal origin predicts the same mass, chemical
composition, and age for both the counter-rotating stellar components.

A proper spectroscopic decomposition that separates the relative
contribution of the counter-rotating stellar components to the
observed galaxy spectrum is therefore needed. Such a decomposition
will allow to disentangle the different formation scenarios by
properly measuring {\it both\/} the kinematics and the stellar
population properties of the counter-rotating stellar disks.
Unfortunately, the techniques that have been developed over the years separated
either the kinematics only (e.g., \citealt{Kuijken+93}) or stellar
populations only (e.g., \citealt{CidFernandes+05, Johnston+12}).

Recently, we have presented a new technique of spectroscopic
decomposition  which we have successfully applied to measure both the
kinematics and stellar population properties of the counter-rotating
stellar disks of NGC~5719 \citep{Coccato+11}. We found that the
counter-rotating stellar component, which rotates in the same
direction of the ionized gas, is younger, less rich in metals, more
$\alpha$-enhanced, and less luminous than the main galaxy disk. We
therefore provided the crucial information that definitely confirms the
gas-accretion scenario for NGC 5719 previously suggested by
\citet{Vergani+07}.

Soon after, other groups developed similar codes. \citet{Katkov+11b}
applied an independent spectroscopic decomposition technique
\citep{Katkov+11a} to NGC 524, finding a stellar disk that
counter-rotates with respect to the galaxy bulge. Both components are
very old (15--20 Gyr); the disk component displays a steep metallicity
gradient, while the bulge a more constant radial profile.  Very
recently, \citet{Johnston+12b} analyzed a long-slit spectrum obtained
along the major axis of NGC~4550 and found that the counter-rotating
stellar component associated to the ionized gas is younger than the
rest of the galaxy.

In this paper, we apply our spectroscopic decomposition technique to
two disk galaxies, NGC~3593 and NGC~4550, which are known to host two
large counter-rotating stellar disks \citep{Rubin+92, Bertola+96}. We
will: i) characterize the morphology, kinematics, mass surface density
distribution, and stellar populations of both the counter-rotating
components, ii) date the formation of the counter-rotating disks, and
iii) probe their formation mechanism.  We adopt a distance to NGC~3593
of 7 Mpc \citep{Wiklind+92} and a distance to NGC~4550 of 16 Mpc
\citep{Graham+99}.
The paper is organized as follows: in Section \ref{sec:observations}
we present the new integral-field spectroscopic observations; in
Section \ref{sec:decomposition} we describe and apply our
spectroscopic decomposition technique to disentangle the two
counter-rotating components; in Section \ref{sec:results} we present
the results; in Section \ref{sec:discussion} we discuss our results
and compare them with the predictions of different formation
scenarios; finally in Section \ref{sec:summary} we present a
conclusive summary.

\begin{figure}
\hbox{
\psfig{file=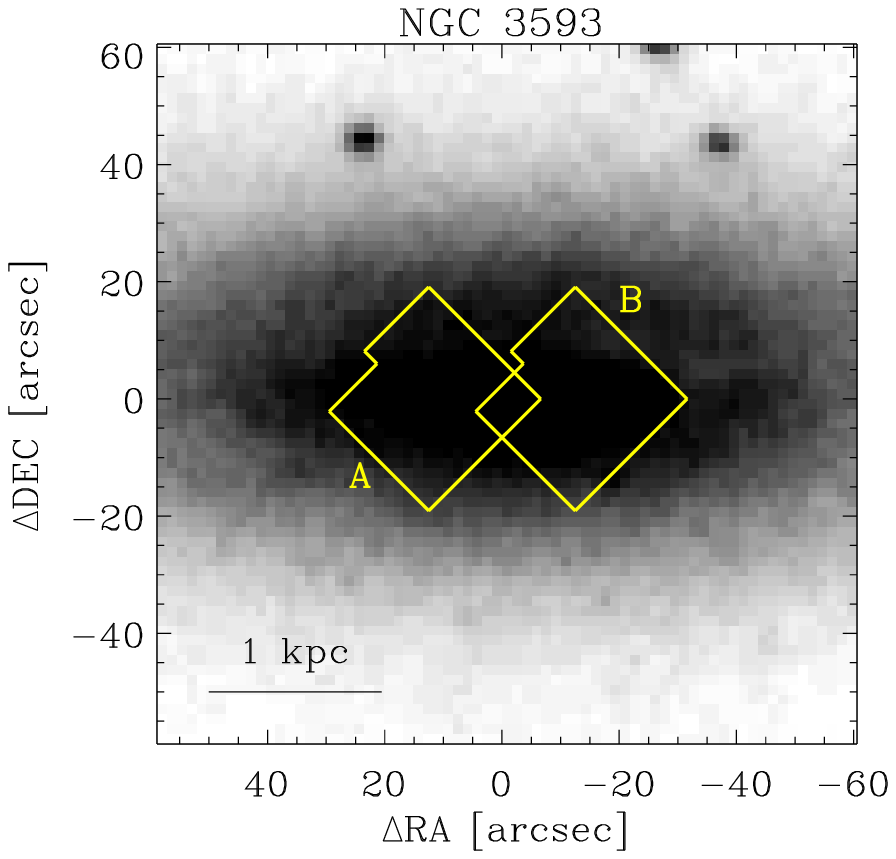,width=4.5cm,clip=}
\psfig{file=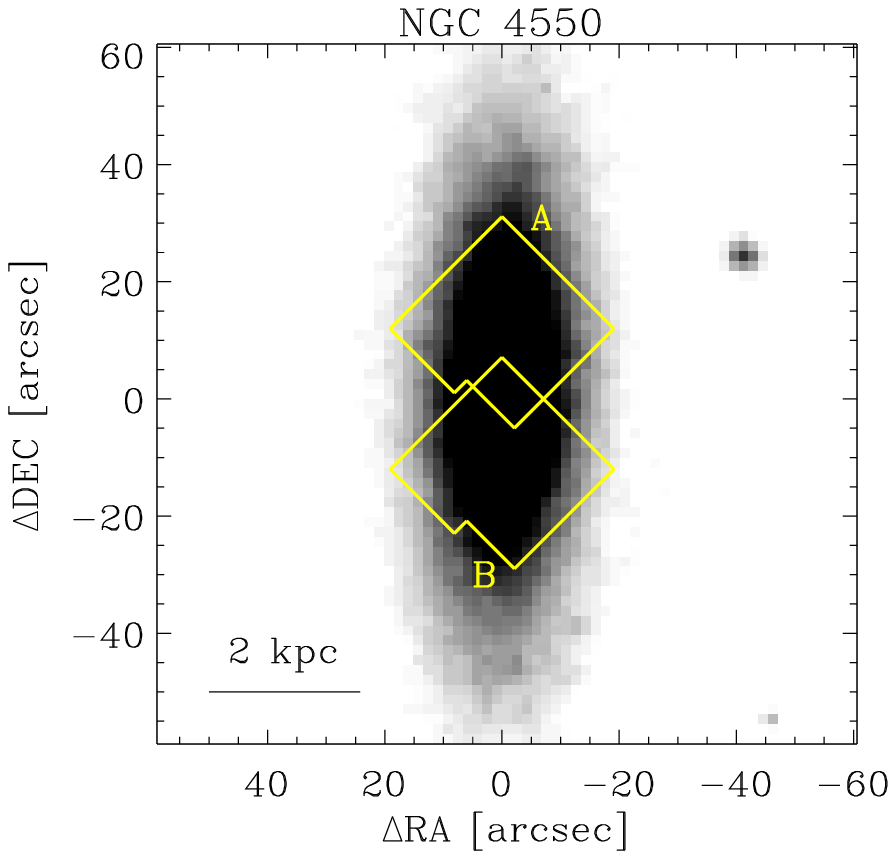,width=4.5cm,clip=}}
\caption{Location of the VIMOS fields of view. The galaxy images are
  from the Digital Sky Survey. North is up and East left. The image
  scale is given on the bottom left corner in each panel.}
\label{fig:fov}
\end{figure}

\section{Observations and data reduction}
\label{sec:observations}   

The integral-field spectroscopic observations were carried out in
service mode with the Very Large Telescope (VLT) at the European
Southern Observatory (ESO) in Paranal during dark time between April
05 and June 02, 2011.  The Unit Telescope 3 was equipped with the
VIsible Multi Object Spectrograph (VIMOS, \citealt{LeFevre+03}) in the
Integral Field Unit (IFU) configuration. We used the 0.67 arcsec
fiber$^{-1}$ resolution and the HR blue grism, covering the spectral
range 4150--6200 \AA\ with a reciprocal dispersion of 0.54
\AA\ pixel$^{-1}$. The instrumental spectral resolution measured at
5200 \AA\ was 2.0 \AA\ (FWHM), equivalent to 115 \kms.  

Observations were organized into different on-target exposures of 2950
s each, alternated to offset exposures of sky fields of 280 s each for
sky subtraction, as detailed in Table \ref{tab:log}. For each galaxy
we observed two overlapping fields-of-view aligned along the galaxy
photometric major axis, at each side of the nucleus. The locations of
the VIMOS fields of view are shown in Figure \ref{fig:fov}.

\begin{table*}
\centering
\caption{Log of the spectroscopic observations.}
\begin{tabular}{l c c c c c c}
\noalign{\smallskip}
\hline
\noalign{\smallskip}                                  
 Target    &  Field  & RA (J2000.0) &  DEC (J2000.0)    & OB        &  Exp. time (+sky)&     Date      \\
           &         &[hh:mm:ss.s]& [dd:mm:ss.s]    &           &    [s]        &  [yyyy-mm-dd] \\
  (1)      &  (2)    &   (3)      &    (4)          &  (5)      &     (6)         &      (7)         \\
 \noalign{\smallskip}                                 
\hline                                                
\noalign{\smallskip}                                  
 NGC~3593 &     A    & 11:15:10.2 &   12:56:53.7     &   540380  &    2850 (+279)  &    2011-05-30 \\
 NGC~3593 &     A    & 11:15:10.1 &   12:56:55.1     &   540382  &    2850 (+280)  &    2011-05-02 \\
 NGC~3593 &     A    & 11:15:10.2 &   12:56:55.8     &   540383  &    2850 (+280)  &    2011-04-05 \\
 NGC~3593 &     A    & 11:15:10.1 &   12:56:54.4     &   540384  &    2850 (+279)  &    2011-05-01 \\
 NGC~3593 &     B    & 11:15:08.6 &   12:56:55.1     &   540385  &    2850 (+279)  &    2011-05-28 \\
 NGC~3593 &     B    & 11:15:08.7 &   12:56:55.7     &   540386  &    2850 (+279)  &    2011-05-28 \\
 NGC~3593 &     B    & 11:15:08.6 &   12:56:54.4     &   540387  &    2850 (+279)  &    2011-05-28 \\
\noalign{\smallskip}                                  
 NGC~4550 &     A    & 12:36:02.6 &   12:05:32.5     &   540388  &    2850 (+279)  &    2011-05-29 \\
 NGC~4550 &     A    & 12:36:02.5 &   12:05:31.9     &   540390  &    2850 (+279)  &    2011-05-29 \\
 NGC~4550 &     A    & 12:36:02.6 &   12:05:33.2     &   540391  &    2850 (+279)  &    2011-05-30 \\
 NGC~4550 &     B    & 12:36:02.7 &   12:05:10.5     &   540392  &    2850 (+279)  &    2011-05-30 \\
 NGC~4550 &     B    & 12:36:02.7 &   12:05:09.9     &   540393  &    2850 (+279)  &    2011-06-02 \\
 NGC~4550 &     B    & 12:36:02.7 &   12:05:11.2     &   540394  &    2850 (+279)  &    2011-06-02 \\
\hline
\noalign{\smallskip}
\label{tab:log}
\end{tabular}
\begin{minipage}{18 cm}
Notes -- 
Col.    1: target name;
Col.    2: VIMOS field-of-view identifier, as given in Fig. \ref{fig:fov};
Cols. 3, 4: coordinates of the center of the field-of-view;
Col.    5: observing block (OB) identifier;
Col.    6: total exposure time on target (and on offset position for sky background 
           evaluation); and
Col.    7: date in which the OB was executed.
\end{minipage}
\end{table*}

Data reduction (bias subtraction, fiber identification and tracing,
flat fielding, wavelength calibration, and correction for instrument
transmission) was performed using the VIMOS ESO pipeline version 2.6.3
under the EsoRex environment\footnote{The VIMOS data reduction
  pipeline and the ESO Recipe Execution Tool are available at
  http://www.eso.org/sci/software/pipelines/.}, plus some {\it ad hoc}
IDL\footnote{Interactive Data Language is distributed by ITT Visual
  Information Solutions. It is available from http://www.ittvis.com/.}
and IRAF\footnote{Image Reduction and Analysis Facility is distributed
  by the National Optical Astronomy Observatories, which are operated
  by the Association of Universities for Research in Astronomy, Inc.,
  under cooperative agreement with the National Science Foundation.}
scripts to correct fiber identification mismatch and remove cosmic
rays.

The different relative transmission of the VIMOS quadrants was
corrected by comparing the intensity of the night-sky emission
lines. The sky offset observations were used to construct sky spectra.
To compensate for the time variation of the relative intensity of the
night-sky emission lines, we compared the fluxes of the sky lines
measured in the offset and on-target exposures. The corrected sky
spectra were then subtracted from the corresponding on-target
exposures. The (very) small amplitude of fringing patterns in the
reduced VIMOS spectra was corrected following the prescriptions by
\citet{Lagerholm+12}.

Each exposure was organized in a data cube using the tabulated
correspondence between each fiber and its position in the field of
view. For each galaxy, the sky-subtracted data cubes were aligned
using the bright galaxy nucleus as reference and co-added into a
single data cube.
Spectra from fibers mapping adjacent regions in the sky were added
together using the Voronoi binning method \citep{Cappellari+03} to
increase the signal-to-noise ratio ($S/N$). Some of the spatial bins in
NGC~3593 were modified to include only spectra from the regions
associated with intense ionized-gas emission or the dust lanes crossing
the galactic disk.  We tested the robustness of our results using
different binning schemes.

\begin{figure*}
\begin{center}
 \hbox{
 \psfig{file=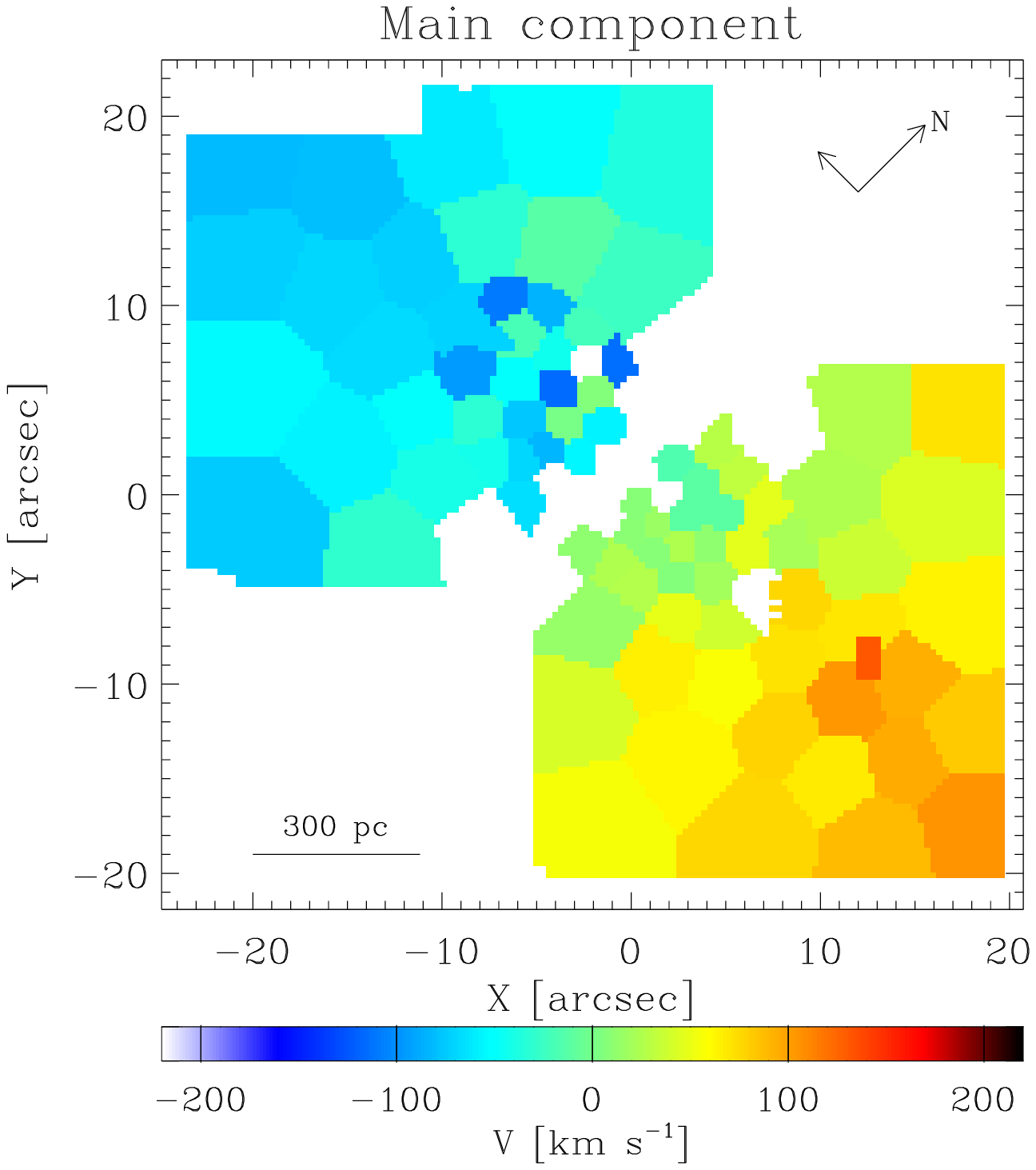,width=6.1cm,clip=,bb=42 360 450 785}
 \psfig{file=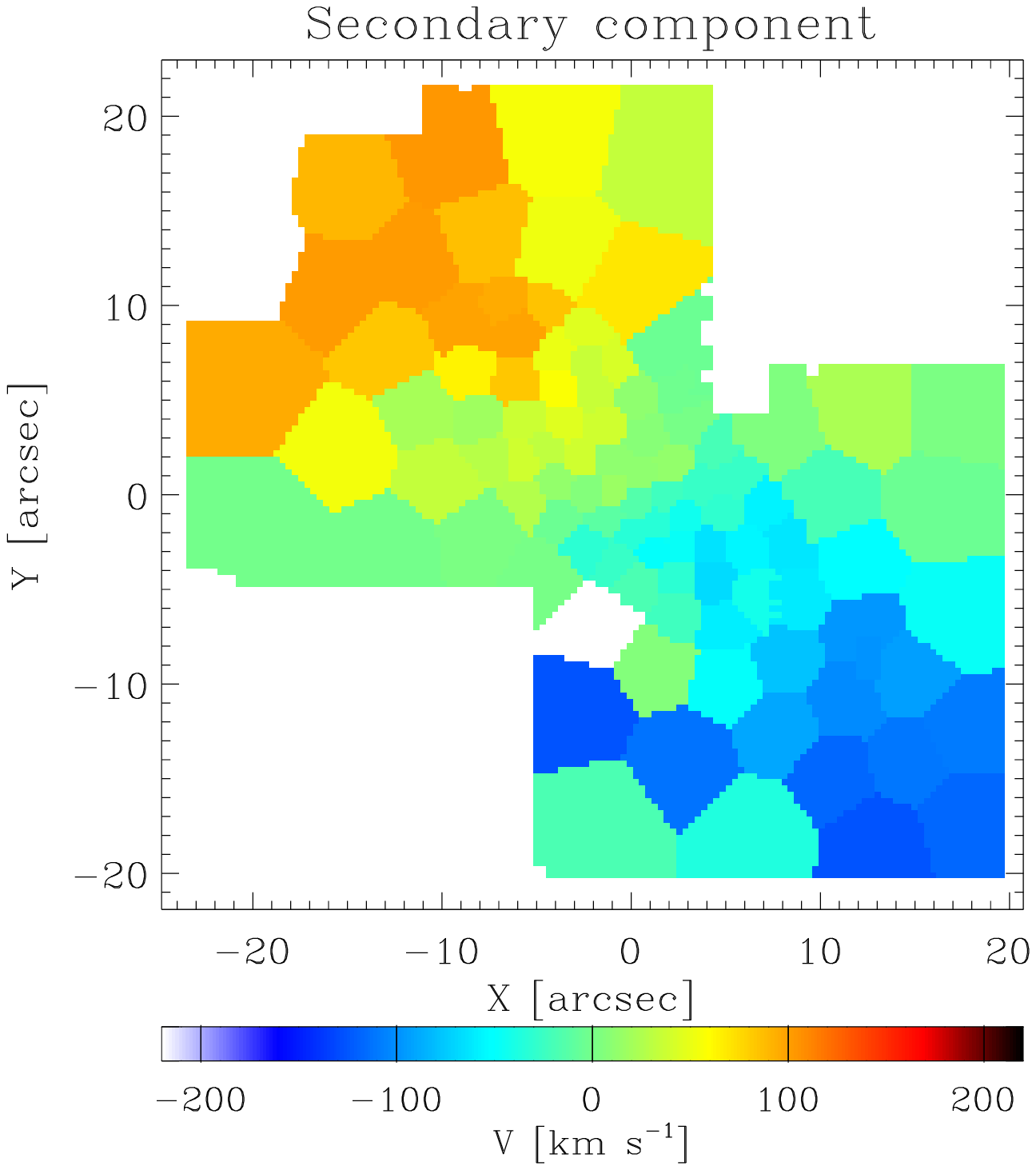,width=6.1cm,clip=,bb=42 360 450 785}
 \psfig{file=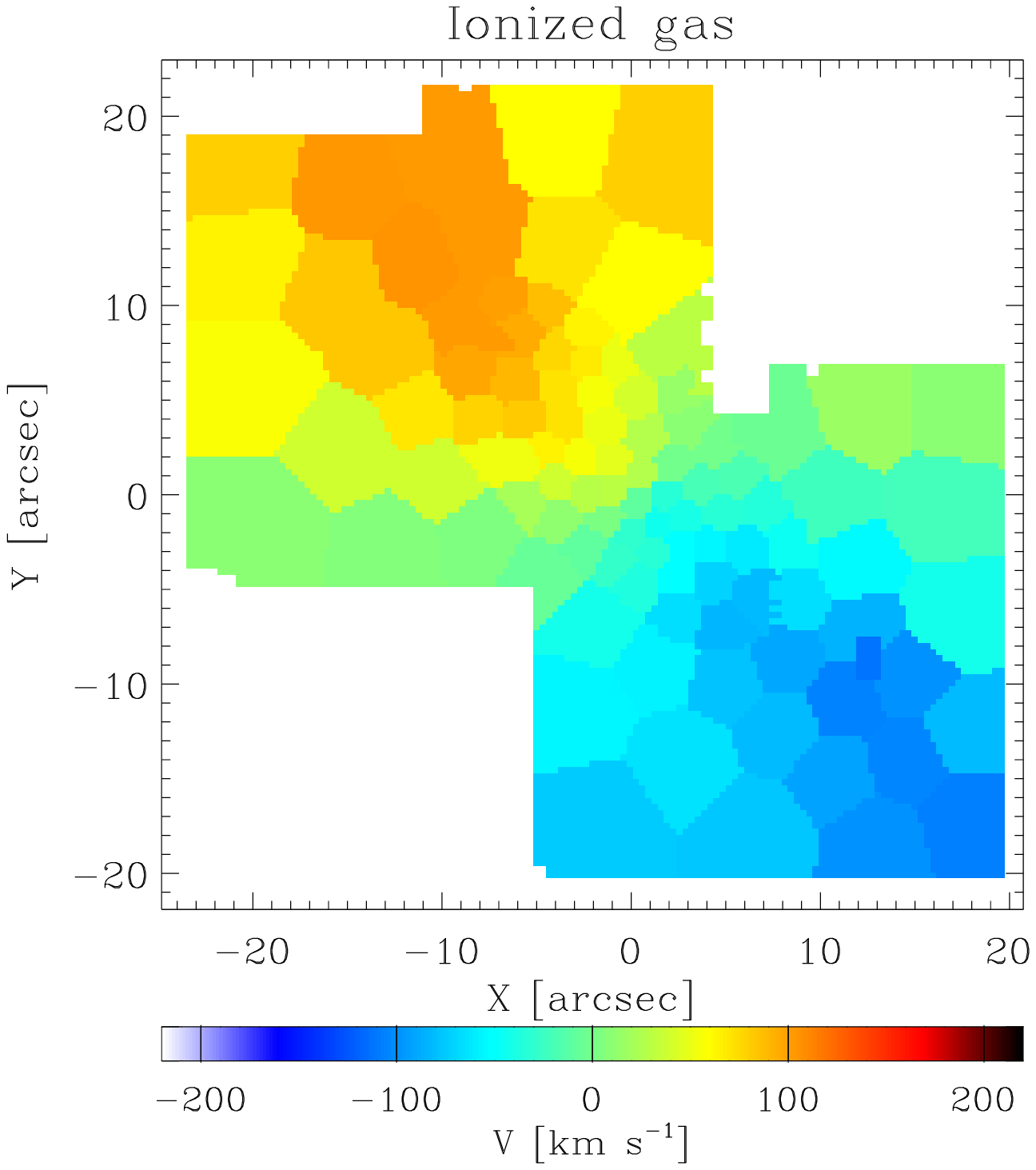,width=6.1cm,clip=,bb=42 360 450 785}} 
\caption{Two-dimensional velocity fields of the main stellar
  component (left panel), secondary stellar component (central panel)
  and ionized-gas component (right panel) in NGC~3593. Scale and
  orientation are given in the left panel. Centers are in RA
  = 11:14:37.0 (J2000.0), Dec. = +12:49:04 (J2000.0).}
\label{fig:kin3593}
\end{center}
\end{figure*}

\begin{figure*}
\begin{center}
 \hbox{
 \psfig{file=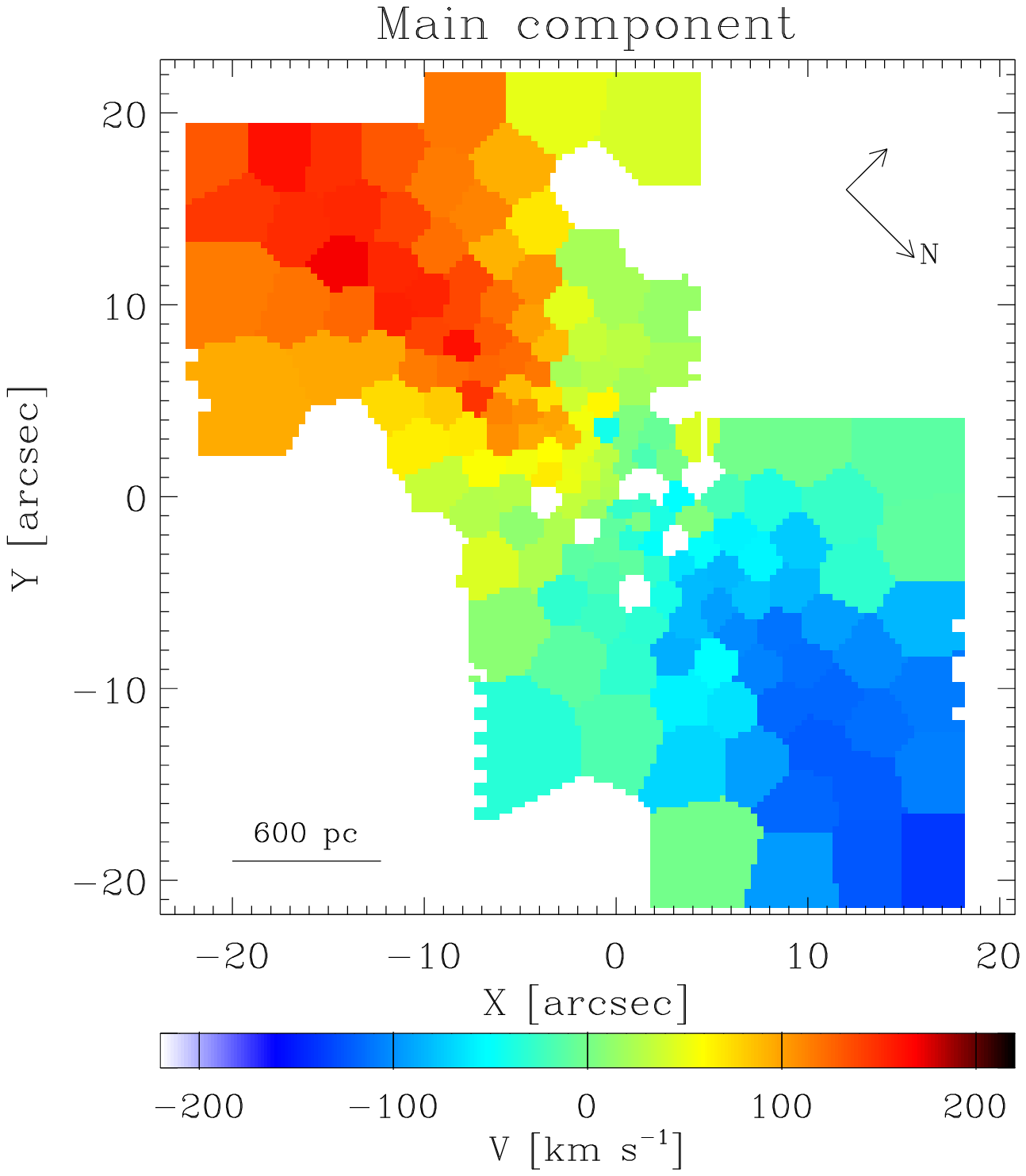,width=6.1cm,clip=,bb=42 360 450 785}
 \psfig{file=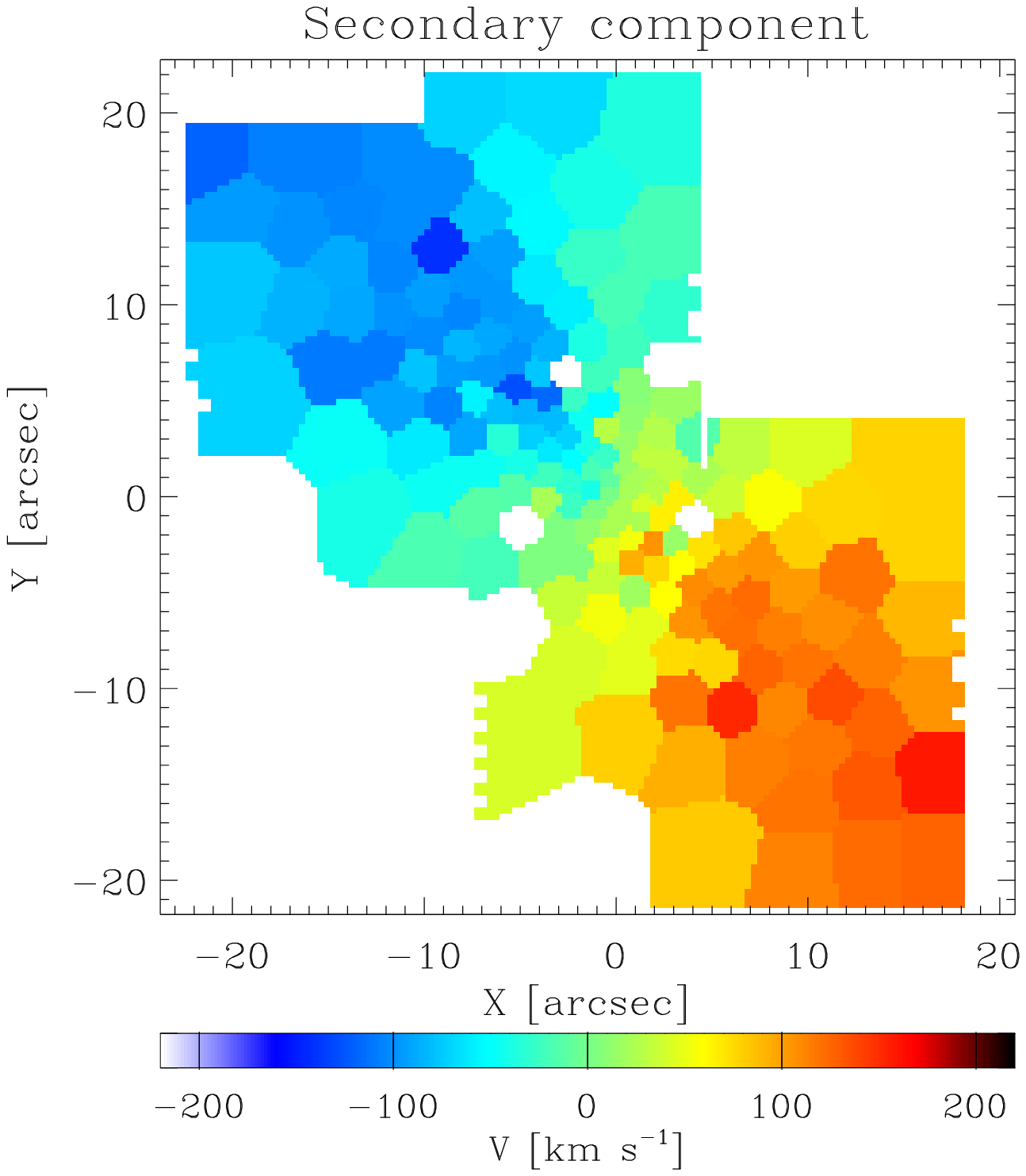,width=6.1cm,clip=,bb=42 360 450 785}
 \psfig{file=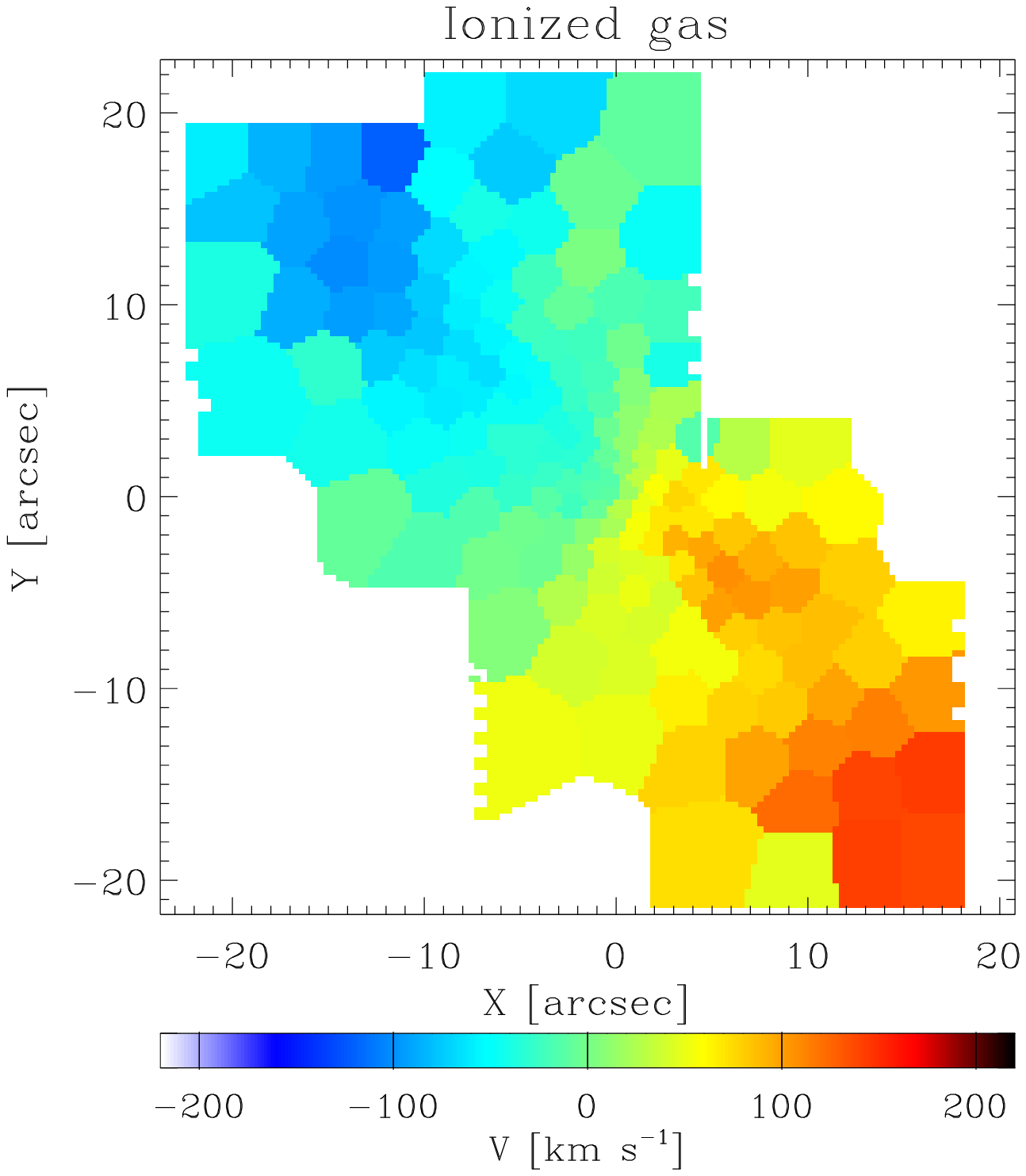,width=6.1cm,clip=,bb=42 360 450 785}} 
\caption{Same as Figure \ref{fig:kin3593}, but for NGC~4550. Centers are in RA
  = 12:35:30.6 (J2000.0), Dec. = +12:13:15 (J2000.0).}
\label{fig:kin4550}
\end{center}
\end{figure*}

\section{Spectroscopic decomposition}
\label{sec:decomposition}

For both galaxies, we separated the contribution of the
ionized-gas component and the two counter-rotating stellar disks to the
observed spectrum using the novel implementation of the penalized
pixel fitting code (pPXF, \citealt{Cappellari+04}) developed in
\citet{Coccato+11}.

In each spatial bin, the code builds two synthetic templates (one for
each stellar component) as linear combination of stellar spectra from
the MILES library \citep{Sanchez+06} at ${\rm FWHM} = 2.54$
\AA\ spectral resolution \citep{Beifiori+11} and convolves them with
two Gaussian line-of-sight velocity distributions (LOSVDs) with
different kinematics. 
Spectra are normalized to their continuum level at 5100 \AA, therefore
the relative contribution of each component to the total spectrum is
in terms of light.
Gaussian functions are added to the convolved synthetic templates to
account for ionized-gas emission lines (\hg, \hb, \oiiipg, and \nipg)
and fit simultaneously to the observed galaxy spectra. Multiplicative
Legendre polynomials are included to match the shape of the galaxy
continuum, and are set to be the same for the two synthetic
templates. The use of multiplicative polynomials also accounts for the
effects of dust extinction and variations of the instrument
transmission.

The spectroscopic decomposition code returns the spectra of two
best-fit synthetic stellar templates and ionized-gas emissions, along
with the best-fitting parameters of luminosity fraction, velocity, and
velocity dispersion. The line strength of the Lick indices of the two
counter-rotating components will be then extracted from the two
best-fit synthetic templates.

\subsection{Errors on measured parameters}

We carried on a set of Monte Carlo simulations on a set of artificial
galaxy spectra to quantify the errors on line-of-sight velocity,
luminosity fraction, and equivalent width as done in
\citet{Coccato+11}. The simulations account for the characteristic
velocity dispersion of our galaxies ($\lesssim90$ \kms), and the
typical $S/N$ ratios, wavelength coverage, and spectral resolution of
our VIMOS observations. Errors in the measured quantities depend on
the quantities themselves, and on their combinations. 

In the nuclear regions where the $S/N$ is high ($S/N \gtrsim 50$
\AA$^{-1}$), the errors are mainly affected by the small velocity
difference of the two stellar components ($\Delta V \lesssim  50$
\kms). In the outermost regions, where the velocity difference between
the two stellar components is large ($\Delta V \gtrsim  200$ \kms),
errors are due to the low $S/N$ ($20 \lesssim S/N \lesssim 30$
\AA$^{-1}$). Errors on the line-of-sight velocity exceed 50 \kms\
 for a luminosity fraction smaller than 20 per cent. Therefore, we
set such a value as a lower limit for a component to be detected.
In NGC~3593 errors in radial velocity do not exceed 40 \kms,
whereas in NGC~4550 they do not exceed 30 \kms. Typical errors in
the equivalent width are 0.6 \AA\ in NGC~3593 and 0.2
\AA\ in NGC 4550.

\section{Results of the spectroscopic decomposition}
\label{sec:results}

We applied our spectroscopic decomposition technique to recover 
the kinematics, surface brightness, properties of the stellar
populations, mass surface density of the counter-rotating stellar
disks, and distribution and metallicity of the ionized-gas component of
NGC~3593 and NGC~4550.

In the following, we present our results. For each stellar component,
we discarded from the analysis all the spatial bins where the
contribution to the total luminosity is lower than either 20 per cent
or $3 \sigma_{\rm noise}$ where the noise standard deviation was
computed from the residuals between measured and best-fitting spectra
in the wavelength range 5000--5400 \AA.

The data extend out to $\sim 25''$ along the photometric major
axis of the two galaxies, which correspond to $\sim$0.8 kpc and
$\sim$2 kpc for NGC~3593 and NGC~4550, respectively.  The spatial
sampling ranges from $\sim 1.3\times1.3$ arcsec$^2$ in the central
regions up to $\sim 8\times8$ arcsec$^2$ in the outermost bins.

From now on, we will refer as ``main'' to the most massive stellar
component, and ``secondary'' to the other. In the equations, we will
indicate them with the subscripts $_{\rm 1}$ and $_{\rm 2}$,
respectively. We will show in Section \ref{sec:mass} that in both
galaxies the secondary and less massive component is the one rotating in
the same direction of the ionized gas.

\begin{figure*}
\begin{center}
\vbox{
 \epsfig{file=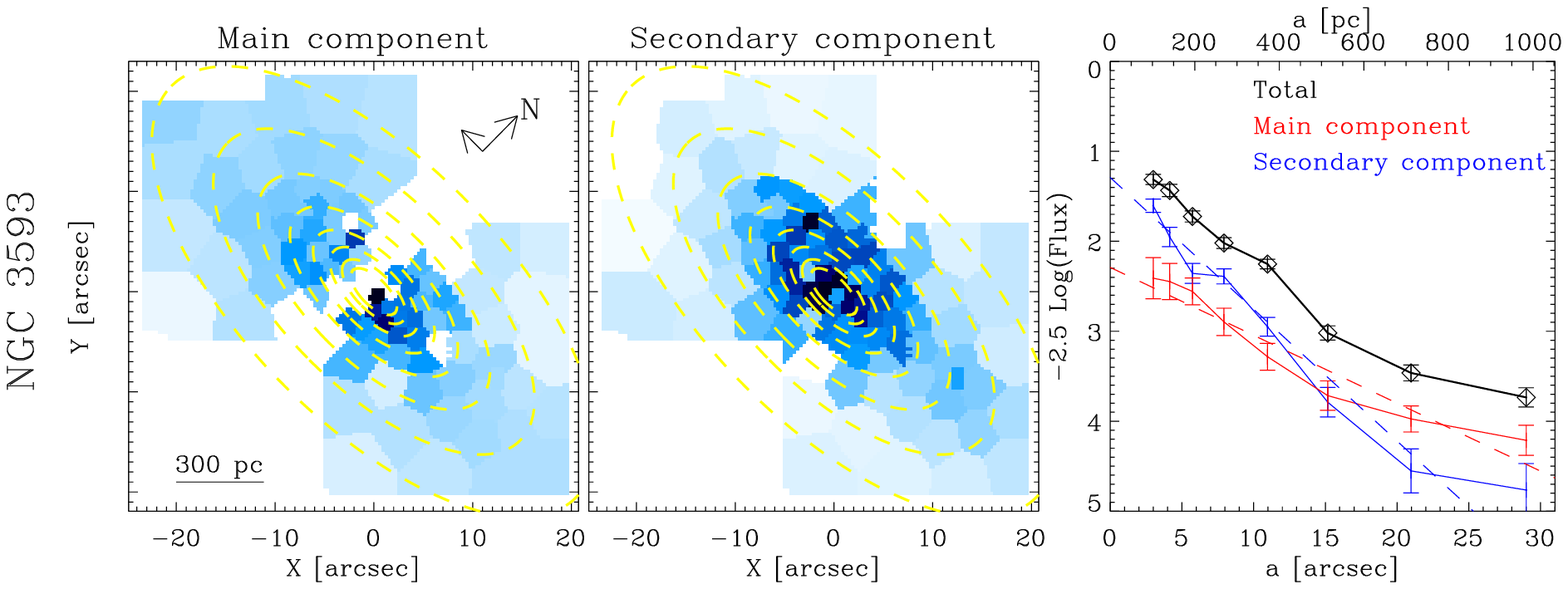,width=15.cm, bb= 14 350 559 560, clip=} \\
 \epsfig{file=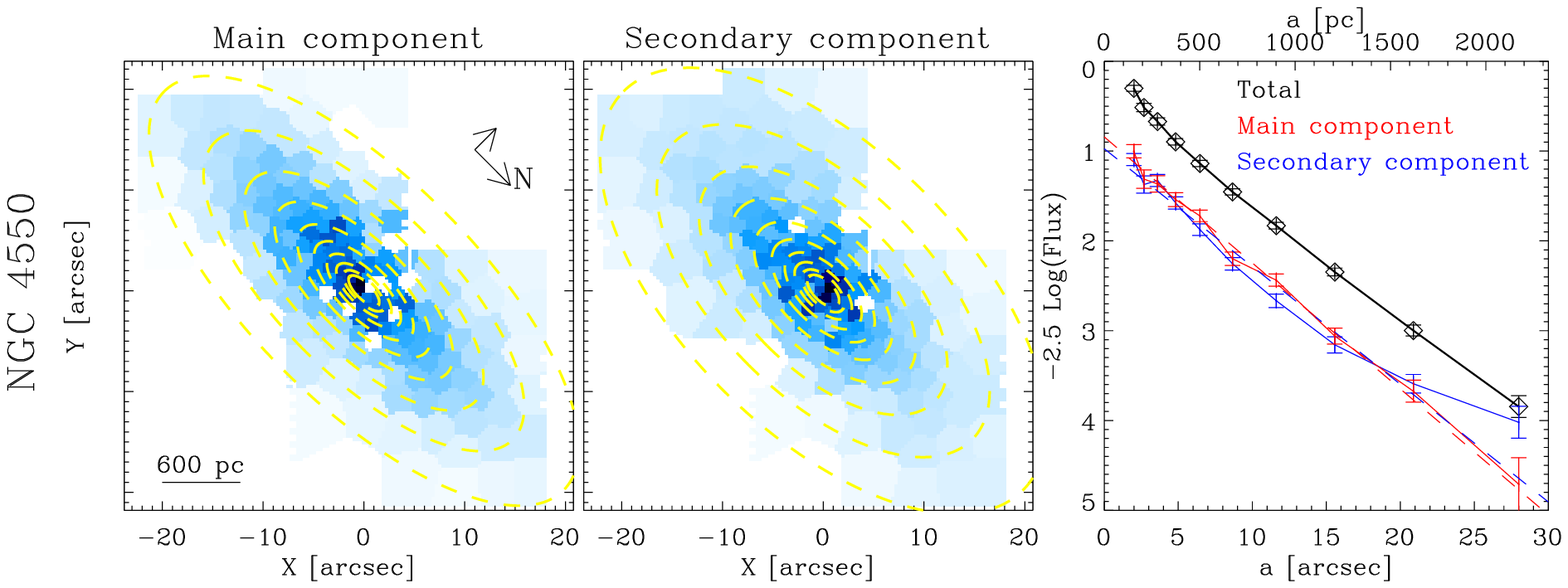,width=15.cm, bb= 14 350 559 575, clip=}}
\caption{Maps (left and central panel) and radial profiles (right
  panels) of the surface brightness of the main and the secondary
  stellar components of NGC~3593 (upper panels) and NGC~4550 (lower
  panels). Scale and orientation of the maps are given in the left
  panels. The black, red, and blue solid lines correspond to the total
  surface brightness and to the surface brightness of the main and the
  secondary stellar components, respectively. The dashed red and
  dashed blue lines correspond to the best-fitting exponential disks
  to the surface brightness of the main and secondary stellar
  components, respectively.  Dashed ellipses represent the boundaries
  of the elliptical annuli where the median surface brightnesses were
  computed.}
\label{fig:surface}
\end{center}
\end{figure*}

\subsection{Stellar and ionized-gas kinematics}
\label{sec:kinematics}

The velocity fields of the main and secondary stellar
components and ionized gas of NGC~3593 and NGC~4550 are shown in
Figures \ref{fig:kin3593} and \ref{fig:kin4550}, respectively.

In the case of NGC~4550, the $S/N$ was high enough to allow the two
LOSVDs to have different velocity dispersions.  The $S/N$ is lower in
the case of NGC~3593, and therefore we decided to decrease the degree
of freedom in the fit by setting the two LOSVDs to have the same
velocity dispersion in the majority of the bins. This is a good
approximation if we take into account the small difference in velocity
dispersion between the two stellar components measured by
\citet{Bertola+96} on long-slit spectra\footnote{\citet{Bertola+96}
  measured $\sigma\simeq 60$ \kms\ in the center and $\simeq 70$ \kms\ at
  $30''$ where the main and the secondary stellar components dominate,
  respectively.}, the spectral resolution, and the mean instrumental
velocity sampling ($\simeq 30$ \kms\ pixel$^{-1}$ at 5100 \AA) of the
VIMOS observations.

In both galaxies we did not observe a central peak in the stellar
velocity dispersion; this indicates that the contribution of the bulge
component is negligible.

Our results confirm the presence in both galaxies of a stellar disk
and a ionized-gas disk that counter-rotate with respect to the main
galaxy disk.

\begin{figure*}
\begin{center}
\vbox{
 \psfig{file=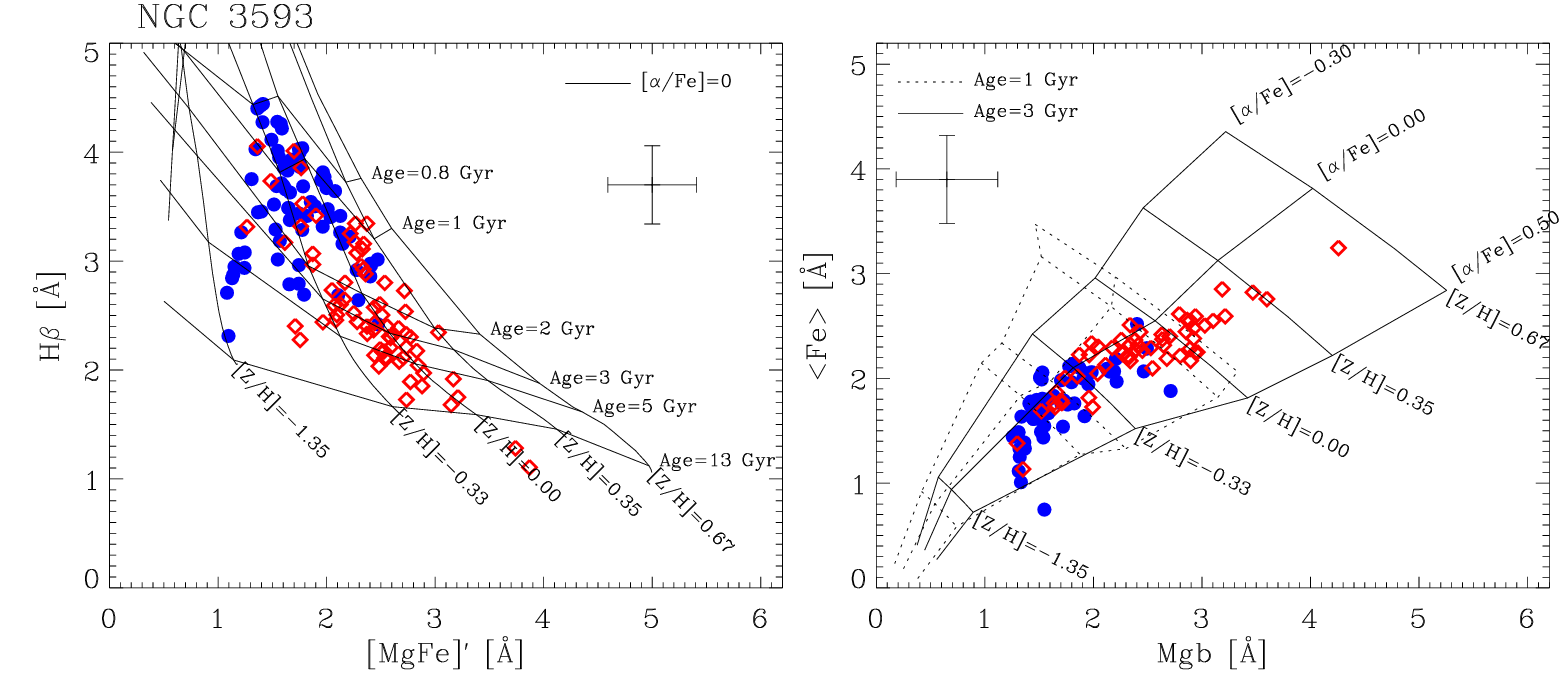,width=16.4cm,clip=, bb= 68 52 764 360}
 \psfig{file=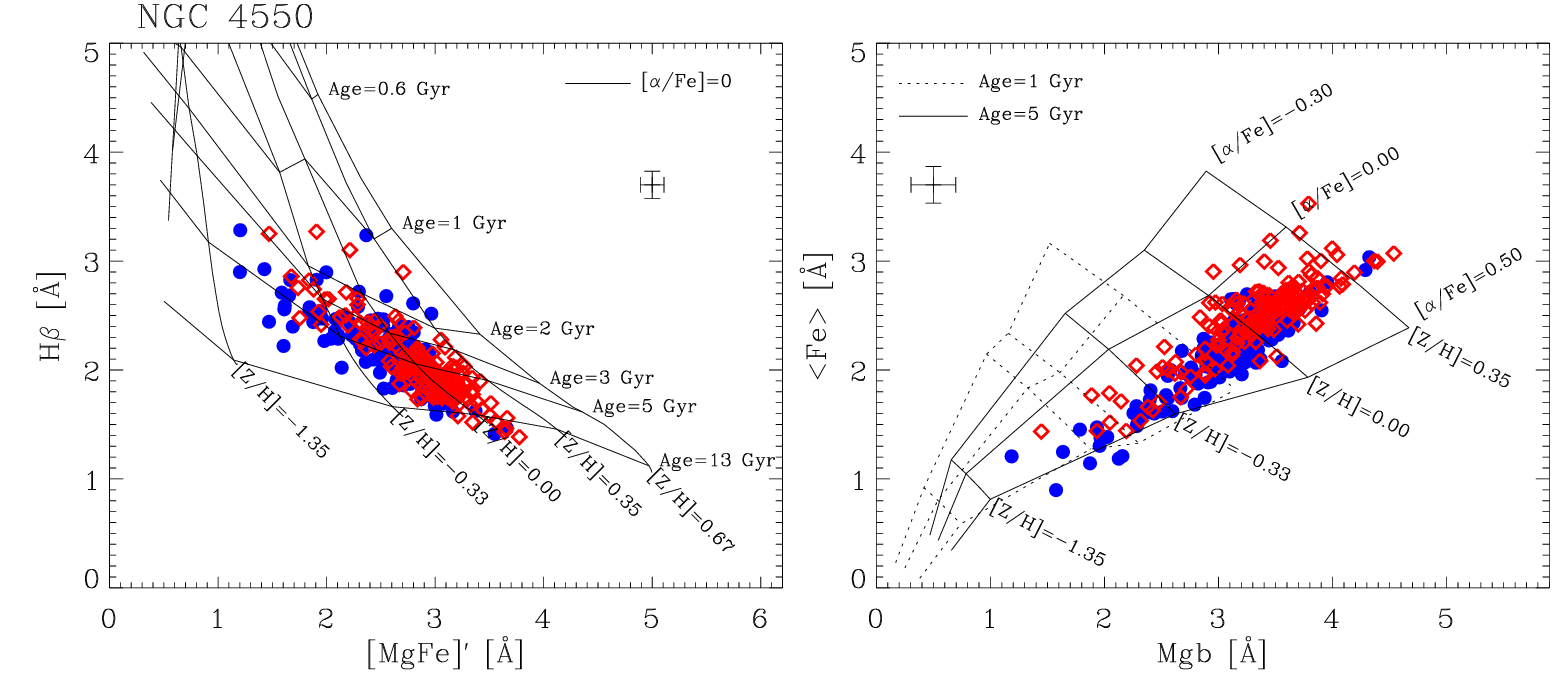,width=16.4cm,clip=, bb= 68 52 764 360}
}
\caption{Equivalent width of the Lick indices in the main (red
  diamonds) and secondary (blue circles) stellar components in NGC~3593
  (upper panels) and NGC~4550 (lower panels). Predictions from single
  stellar population models by \citet{Thomas+11} are
  superimposed. Crosses indicate mean error bars associated to the
  equivalent widths.}
\label{fig:indices}
\end{center}
\end{figure*}

\bigskip
\noindent{\bf \it NGC 3593}

The main stellar component reaches a line-of-sight rotation of $V_1
\simeq 80$ \kms\ at $R\simeq 20''$ ($\simeq 700$ pc) along a direction
consistent with the galaxy photometric major axis. This rotation is
slightly higher than that measured by \citet[$V \simeq 70$
\kms]{Bertola+96} at the same distance. This is because in the
previous work the line of sight velocity distributions of the two
stellar components were not fully resolved at $20''$. The secondary
stellar disk and ionized-gas components counter-rotate with respect to
the main component, and their velocity is $V_2 \simeq 100$ \kms,
consistent with that of the ionized-gas measured by
\citet{Bertola+96}.

The stellar velocity dispersion ranges between $30 \lesssim \sigma
\lesssim 80$ \kms; the velocity dispersion of the ionized gas is $\lesssim
30$ \kms.

\bigskip
\noindent{\bf \it NGC 4550}

The main stellar disk rotates by $V_1 \simeq 140$ \kms\ at $R \simeq
25''$ ($\simeq 2$ kpc), while the secondary stellar disk and the
ionized-gas component rotate slightly slower ($V_2 \simeq 110$
\kms). Our measurements are consistent with the  long-slit
  absorption-line and \oiii\ kinematics of \citet{Rix+92} and
  \citet{Johnston+12b}, and the ionized-gas kinematics of
\citet{Sarzi+06} from \hb\ and \oiii. On the contrary,
\citet{Rubin+92} measured a larger rotation ($V \sim 140$ \kms) for
the \ha, \nii\, and \sii\ emission lines. Unfortunately, we do not have
enough spectral coverage to investigate the difference in the
kinematics of these emission lines.

The velocity dispersions of the two stellar disks are very similar: the
main component has $60 \lesssim \sigma_1 \lesssim 80$ \kms\ in the
center, followed by a decline outside $R>14''$ (1 kpc), where $
\sigma_1 \lesssim 30$ \kms. The velocity dispersion of the secondary
component instead ranges between $40 \lesssim \sigma_2 \lesssim 80$
\kms.  The velocity dispersion of the ionized gas is $\simeq 50$ \kms.

\subsection{Stellar surface brightness}
\label{sec:surfbr}

We extracted the surface-brightness maps of the two stellar disks from
their relative luminosity fraction (obtained in the spectral
decomposition) and the observed total flux (obtained by collapsing the
data-cube along the dispersion direction).

The stellar surface-brightness maps and radial profiles are shown in Figure
\ref{fig:surface}.

\bigskip
\noindent{\bf \it NGC 3593}

In NGC~3593, a preliminary  fit with the IRAF task ELLIPSE
\citep{Jedrzejewski87} indicated that the two stellar disks have the
same mean ellipticity and orientation on the sky within the errors. We
therefore computed their median surface brightnesses along concentric
ellipses, adopting the same ellipticity $\epsilon =0.55$ and position
angle PA$= 90^{\circ}$, both constant with radius.

The surface-brightness profiles of the two stellar disks in NGC~3593
did not reveal any remarkable structure such as rings or spiral arms
above the noise level (Fig. \ref{fig:surface}).
We parametrized their light distributions with two exponential disks
\citep{Freeman+70}, which we found to have different scale lengths:
$h_{\rm 1}=14" \pm 2"$ ($480 \pm 70$ pc) and $h_{\rm 2}=7" \pm 1"$
($240 \pm 35$ pc). \citet{Bertola+96} reported similar scale length
for the secondary component ($h_{\rm 2}=10"$, 340 pc), but larger value for
the main component ($h_{\rm 1}=40"$, 1.4 kpc). The discrepancy is due to the
fact that we probe a different band and our data probe only the inner
$20''$, where the main stellar component is still faint. The different
scale lengths of the counter-rotating stellar disks make the
secondary stellar component to dominate the total galaxy light profile
inside $\simeq 15"$, whereas the main stellar disk dominates
outside. The transition radius between the light contribution of the
two stellar disks is consistent with the photometric decomposition by
\citet{Bertola+96}.

\begin{figure*}
\begin{center}
\vbox{
 \psfig{file=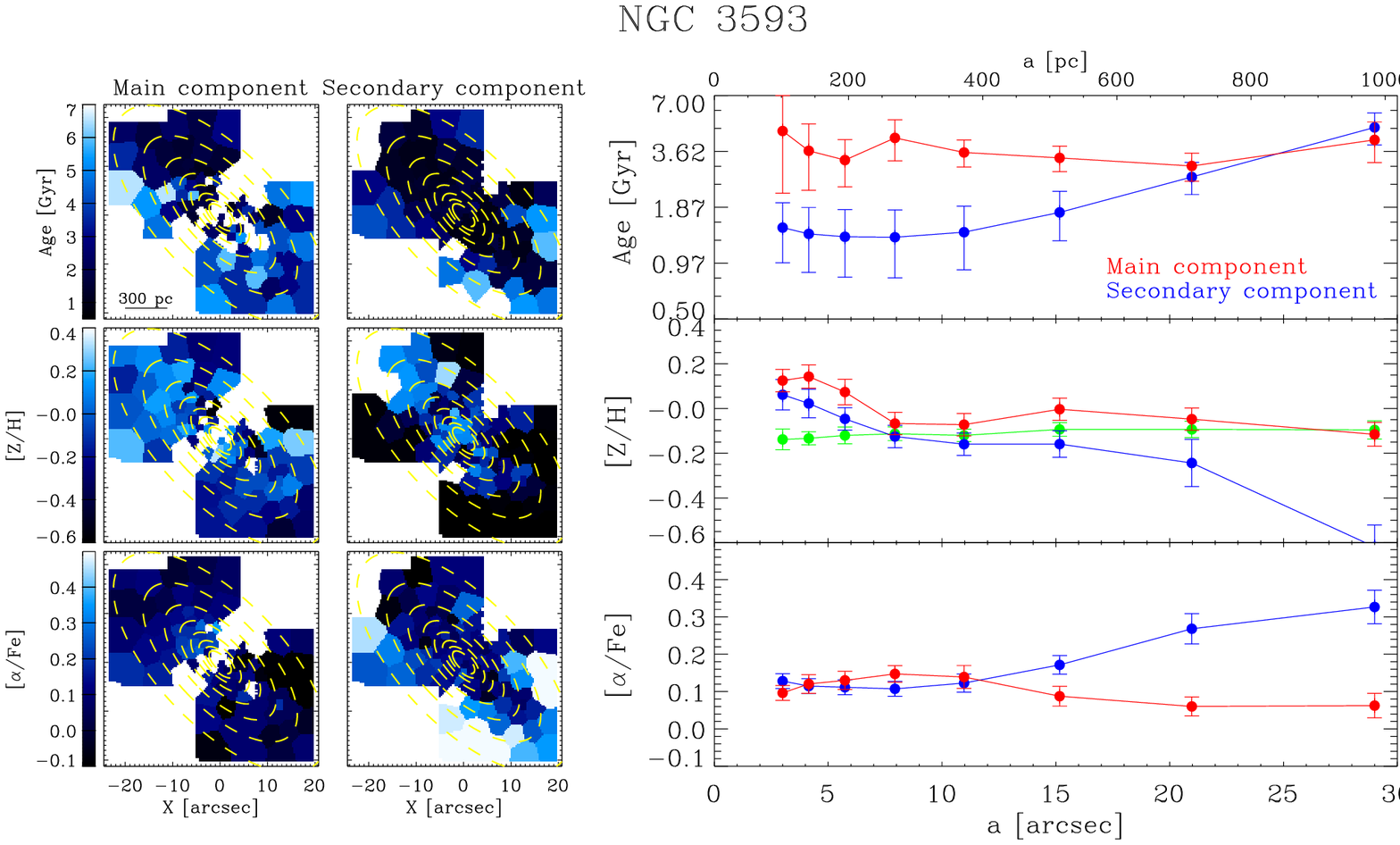,  clip=, width=15.cm,bb=15 360 620 720}\\ 
 \psfig{file=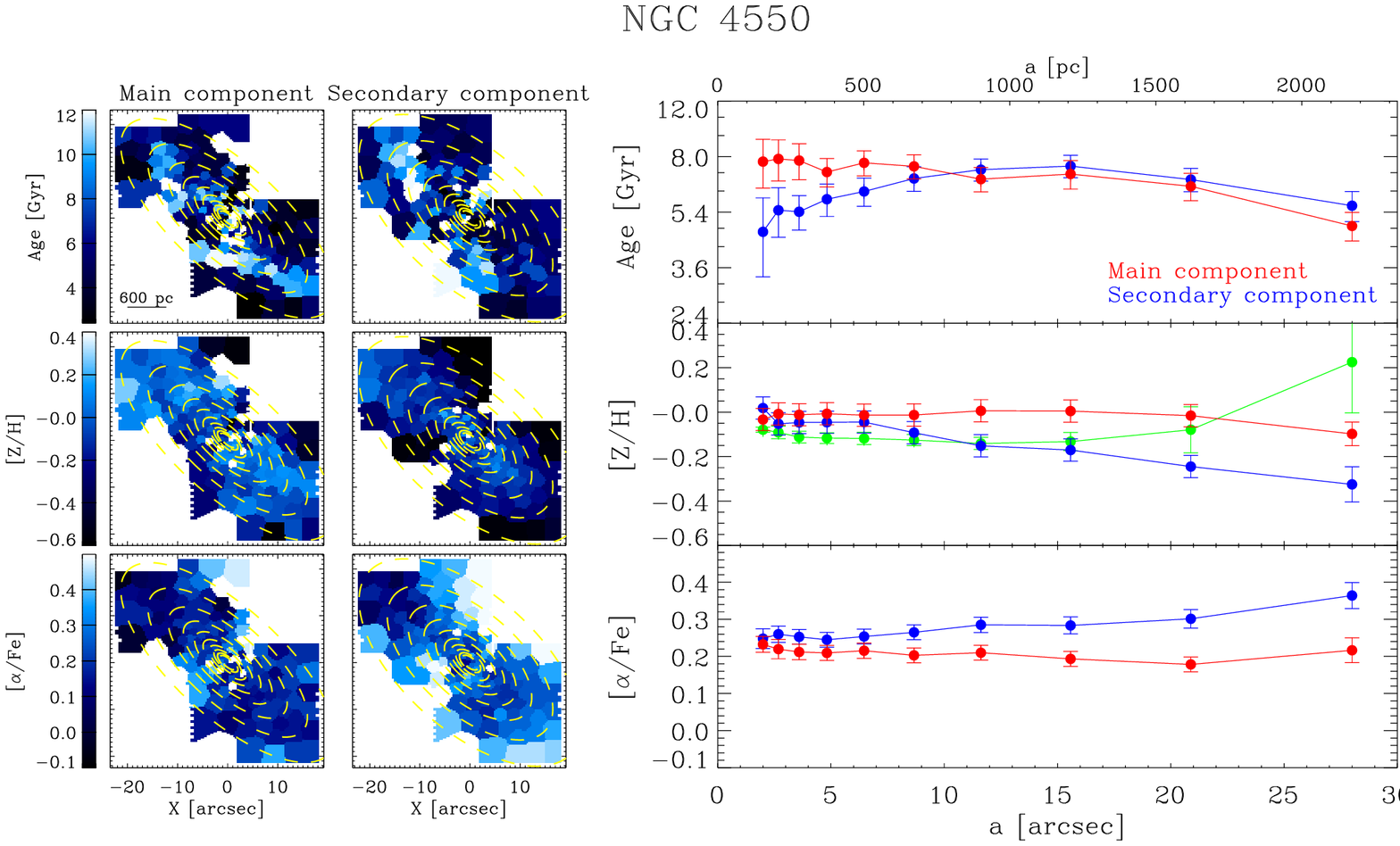,  clip=, width=15.cm,bb=15 360 620 740}
}
\caption{Stellar population parameters in NGC 3593 (upper panels) and
  NGC 4550 (lower panels). Left panels: maps of the age (top), metallicity
  (middle) and $\alpha$-enhancement (bottom) of the two stellar
  counter-rotating components, as derived by fitting the
  \citet{Thomas+11} models to the measured Lick indices. Dashed
  ellipses show the elliptical annuli used to bin the data and extract
  the radial profiles shown on the right panels. Orientation is the
  same as in Figures \ref{fig:kin3593} and \ref{fig:kin4550}. Right
  panels: radial profiles of the total luminosity-weighted values
  within concentric ellipses of the age (top), metallicity (middle),
  and $\alpha$-enhancement (bottom). Red and blue lines represent the
  main and secondary stellar components, respectively. The green line
  in the [Z/H] panels represents the metallicity of the ionized-gas
  component. }
\label{fig:rad_ssp}
\end{center}
\end{figure*}

\bigskip
\noindent{\bf \it NGC 4550}

In NGC~4550, a preliminary fit with the IRAF task ELLIPSE indicates that the
two stellar disks have similar orientation on the sky but different
mean ellipticity. We therefore compute the median surface brightness
along concentric ellipses with the same PA$=0$ and different
ellipticities ($\epsilon_{\rm 1}=0.6$, $\epsilon_{\rm 2}=0.5$).

The two stellar disks in NGC~4550 have the same scale length within
errorbars ($h_{\rm 1}=7\farcs7 \pm 0\farcs4$, i.e. $600 \pm 30$ pc,
and $h_{\rm 2}=8\farcs3 \pm 0\farcs7$, i.e. $650 \pm 50$ pc) and the
same luminosity, as found by \citet{Rix+92} from long-slit
data. Their apparent flattening is different, meaning that they have
different scale heights (Fig. \ref{fig:surface}).  Assuming the
observed mean ellipticities ($\epsilon_{\rm 1}= 0.6$ and $\epsilon_{\rm 2}= 0.5$,
see Sect. \ref{sec:surfbr}) and a common inclination for the two
stellar disks, we find that the minimum allowed inclination is
$i=67^{\circ}$, which leads to intrinsic flattenings of $q_{0, \rm 1}=
0$ and $q_{0, \rm 2}= 0.32$. The intrinsic flattening $q_0$ was
computed from $q_0^2 (1 - \cos^2{i}) = q^2 - \cos^2{i}$, for an oblate
spheroid with measured flattening $q = 1 - \epsilon$ and inclination
$i$.

\citet{Emsellem+04} presented the mean stellar velocity field showing
that the rotation is dominated by one component (corresponding to our
main component) in a thin region along the major axis, while in a
broad region away from the major axis the rotation is dominated by the
other component (which rotates accordingly to the ionized gas). Their
explanation is that the two components must have very different scale
heights \citep{Cappellari+07}. This is not in contrast with the mild
difference in the apparent flattening we found instead. In fact, the
limitation of a single component fit is that it indicates only where
one component is dominant; this leads to an overestimation of the
flattening difference. On the contrary, our spectroscopic
decomposition allows to measure both the two underlying spatial
surface-brightness distributions, and to properly quantify the
flattening. Alternatively, one can assume the same intrinsic
flattening $q_0$ for both components. This would imply that the two
stellar disks have different inclinations (\citealt{Afanasiev+02}).
Extrapolating the typical value of S0/Sa galaxies ($q_0=0.18$,
\citealt{Guthrie+92}) to NGC 4550 (E7/S0), it results: $i_{\rm 1} =
68^\circ$ and $i_{\rm 2} = 62^\circ$.

\subsection{Line-strength indices and stellar populations}
\label{sec:ssp}

In each galaxy, we derived the properties of the stellar populations
of the two stellar counter-rotating components by measuring the line
strength of the Lick indices \hb, \mgb, \fei, and \feii\
\citep{Worthey+94} on the best fit synthetic templates returned by the
spectroscopic decomposition.
To this aim, the spectra were set to rest-frame by adopting the
measured radial velocity and convolved with a Gaussian function to
match the spectral resolution of the Lick system ($\rm FWHM = 8.4$
\AA, \citealt{Worthey+97}).
From the measured indices, we calculated the mean iron index $\langle
{\rm Fe} \rangle$ = $\left( {\rm \fei} + {\rm \feii} \right)/2$
\citep{Gorgas+90} and the combined magnesium-iron index [MgFe]$' =
\sqrt{{\rm \mgb} \left( 0.82 \cdot {\rm \fei} + 0.28 \cdot {\rm \feii}
  \right)}$. The [MgFe]$'$ index is almost independent from
$\alpha$-enhancement and hence serves best as a metallicity tracer
\citep{Thomas+03}.

In Figure \ref{fig:indices} we compare the measured indices with the
predictions of single-age stellar populations (SSP) model by
\citet{Thomas+11} that takes into account abundance ratios of $\alpha$
elements different from solar.
Alternative line-strength indices as H$\beta_0$ \citep{Cervantes+09}
led to consistent results. We decided to present the analysis based on
the ``classic'' \hb\ spectral index for consistency to our previous
work on NGC 5179 \citep{Coccato+11}.

We then fitted the \citet{Thomas+11} models to the measured indices to
get two-dimensional maps of the luminosity-weighted mean stellar
population age, metallicity ([Z/H]), and $\alpha$-enhancement
([$\alpha$/Fe]) of the two stellar components in the sample
galaxies. They are shown in the left panels of Figure
\ref{fig:rad_ssp}, and their luminosity-weighted values are given in
Table \ref{tab:mean_values}.  We also studied the radial dependence of
these stellar population parameters, by computing their
luminosity-weighted mean values along concentric ellipses using the
same geometrical parameters adopted in Sect. \ref{sec:surfbr}. They
are shown in the right panels of Figure \ref{fig:rad_ssp}. Errorbars
on the radial profiles are computed as $\sigma_{\rm SSP} / \sqrt{N}$,
where $\sigma_{\rm SSP}$ is the standard deviation of the measurements
within each ellipse, and $N$ is the number of Voronoi bins within the
ellipse.

MILES stars belong to the Milky Way and this makes the stellar library
trapped along the essentially one-parameter trend with a rise of
[Mg/Fe] while decreasing [Fe/H] that translates into an
anti-correlation between [$\alpha$/Fe] and [Z/H]. This trend has a
spread of $\sim$ 0.5 dex in [Mg/Fe] and it is steeper in the
metal-poor regime ($\rm [Fe/H]\,<\,-0.4$) while it is shallower for
stars with $\rm [Fe/H]\,>\,-0.4$ (e.g., \citealt{Milone+11}). Since we
measured the Lick indices for linear combinations of MILES stars, the
final values of [$\alpha$/Fe] and [Z/H] we derived from SSP models
could be not completely independent.

The anti-correlation between [$\alpha$/Fe] and [Z/H] is observed in
the outer regions ($R\,>\,15''$) of our galaxies where we measured low
values of metallicity (Fig. \ref{fig:rad_ssp}), whereas it is less
strong in the central regions ($R\,<\,15''$) where the metallicity is
higher ($\rm 0.15\,<\,[Z/H]\,<\,-0.25$). The ability of the fitting
tool of recovering the anti-correlation between $\alpha$-enhancement
and metallicity is an independent indication of its robustness.

We used [$\alpha$/Fe] as a proxy of the star-formation time scale
$\Delta\,t$ following the approximation by \citet[Equation
4]{Thomas+05}. The main sources of errors on $\Delta\,t$ are the
uncertainties about SN Ia progenitors and delay time distributions
(e.g., \citealt{Pritchet+08}) and dependence of our measurements of
$\alpha$-enhancement on metallicity.

\begin{table}
\centering
\caption{Luminosity-weighted values for the stellar population parameters 
of the stellar disks in NGC~3593 and NGC~4550.}
\begin{tabular}{l c c c}
\hline
\noalign{\smallskip}
                &     \mage        &      \mfe         &  \malpha \\
                &    [Gyr]         &                   &                       \\
\noalign{\smallskip}
\hline
\noalign{\smallskip}
NGC~3593        &                  &                    &                     \\
\ \ \ Main:     &   $3.6 \pm 0.6$  &  $-0.04 \pm  0.03$ & $0.09 \pm 0.02$    \\
\ \ \ Secondary:   &   $2.0 \pm 0.5$  &  $-0.15 \pm  0.07$ & $0.18  \pm 0.03$      \\
\noalign{\smallskip}
\noalign{\smallskip}
NGC~4550        &                  &                    &                    \\
\ \ \ Main:     &   $ 6.9 \pm 0.6$ & $-0.01 \pm 0.03$   & $0.20 \pm  0.02$   \\
\ \ \ Secondary:   &   $ 6.5 \pm 0.5$ & $-0.13 \pm 0.04$   & $0.28  \pm  0.02$    \\
\noalign{\smallskip}
\hline 
\noalign{\smallskip}
\end{tabular}
\begin{minipage}{9cm} 
  Notes-- \mfe\ and \malpha\ are given in logarithm of solar
  units. Errors are computed as the standard deviation of the
  measurements divided by the square root of the number of spatial
  bins.
\end{minipage} 
\label{tab:mean_values}
\end{table}

\begin{figure*}
\begin{center}
\vbox{
 \psfig{file=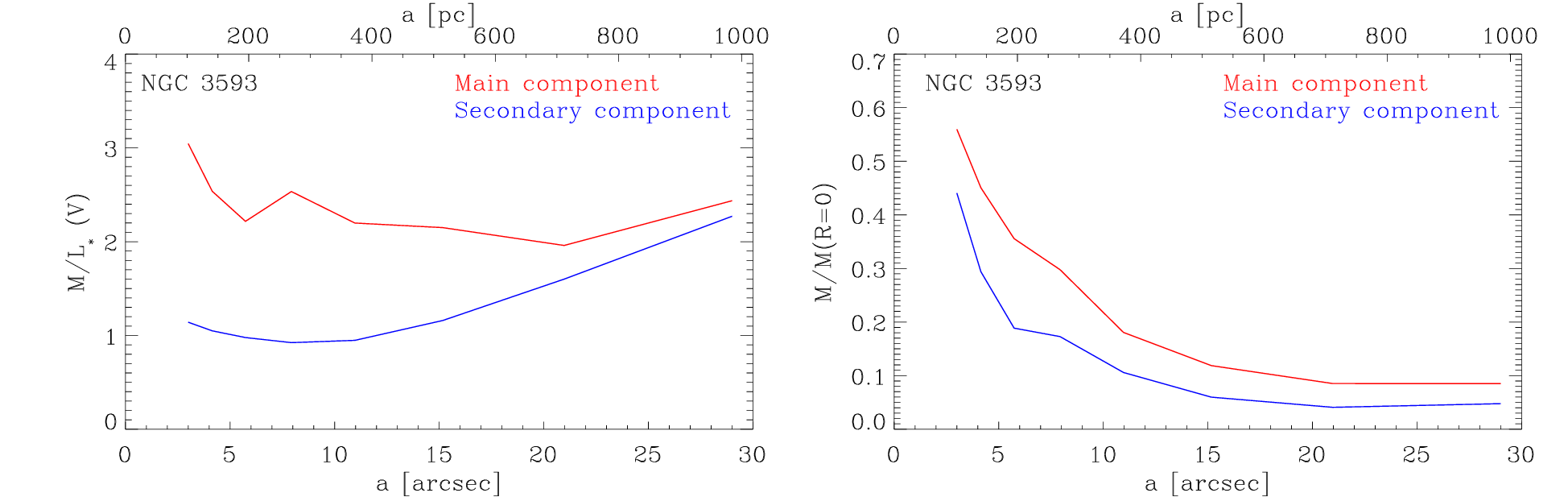,width=18cm,clip=,bb=85 360 980 661}
 \psfig{file=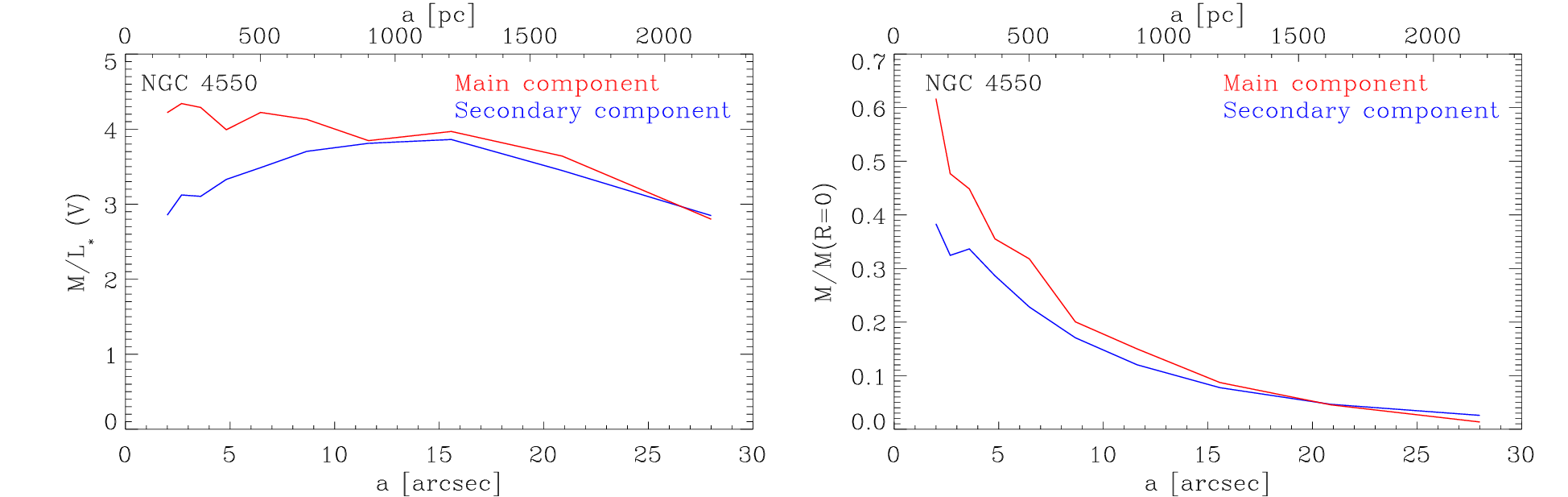,width=18cm,clip=,bb=85 360 980 690}}
\caption{Radial profiles of the mass-to-light ratio in the $V$ band (left
  panels) and mass surface density profiles relative to the central value (right panels)
  for NGC 3593 (upper panels) and NGC 4550 (lower panels). Red and
  blue lines refer to the main and the secondary stellar components,
  respectively.  The x-axis displays the semi-major axis of the ellipses
  shown in Fig. \ref{fig:surface}.}
\label{fig:mass_profile}
\end{center}
\end{figure*}

\bigskip
\noindent{\bf \it NGC 3593} 

The two stellar disks in NGC~3593 have different spectral
properties. This difference is visible in the diagnostic plots of
Fig. \ref{fig:indices}, which compare the equivalent widths of \hb, Mg
and Fe absorption lines.  Unfortunately, the two-dimensional maps of
the stellar population properties (age, [Z/H], [$\alpha$/Fe]) do not
highlight this separation as clearly, because of the scatter in the
measurements; the radial profiles in Fig. \ref{fig:rad_ssp} serve
better to this purpose.  Although in some radial bins the two stellar
components have consistent stellar population properties within the
errorbars, there are clear systematic trends with radius.

The age difference between the two components is clearly visible for
$R<15''$ ($\sim 500$ pc). In this radial range, the main component is
the oldest (\mage$_{\rm 1}(R<15'') = 3.7 \pm 0.6$ Gyr) and the
secondary component, associated with the gas, is the youngest
(\mage$_{\rm 2}(R<15'') = 1.4 \pm 0.2$ Gyr).

The main component is systematically more metal rich (\mfe$_{\rm
  1}(5''<R<15'') =-0.04 \pm 0.04$ dex) than the secondary component
(\mfe$_{\rm 2}(5''<R<15'') =-0.12 \pm 0.03$ dex). Both components have
a metallicity peak in the central $\sim$150 pc: main
component: \mfe$_{\rm 1}(R<5'') =+0.13 \pm 0.04$ dex, secondary
component: \mfe$_{\rm 2}(R<5'') =+0.04 \pm 0.06$ dex. 
The metallicity gradient of the secondary component is steeper
  than that of the main component.

The main and secondary components are both characterized by an
intermediate star-formation time scale ($\Delta t \simeq 2.5$ Gyr) till
$15''$ ($\simeq 500$ pc), whereas the outer parts of the secondary
component are characterized by a much shorter time scale ($\Delta t
\simeq 0.4$ Gyr).

\bigskip
\noindent{\bf \it NGC 4550} 

The spectral properties of the two stellar disks in NGC~4550 are very
similar, although there is an indication that the main component has
slightly larger mean values of \mgb\ and Fe and smaller mean values of
\hb\ than the secondary component (Fig. \ref{fig:indices}).  Our
line-strength measurements of the Lick indices agree with the average
values measured by \citet{Johnston+12b} for both the counter-rotating
components. Nevertheless, \citet{Johnston+12b} derived a mean age of
the secondary component (2.5 Gyr), which is younger than ours (6.5
Gyr). This is due to the differences in the adopted SSP models and
performed analysis. Indeed, we used the models of \citet{Thomas+10}
and computed the luminosity-weighted average of all the SSP parameters
we measured on the two-dimensional field of view, whereas
\citet{Johnston+12b} fit the models of \citet{Vazdekis+10} to the
arithmetic mean of the equivalent widths they measured in long-slit
data.

The total galaxy luminosity-weighted age within $1R_e \simeq
15\farcs5$ (1.2 kpc) is \mage$ = 6.9 \pm 0.4$ Gyr. This is consistent
with the luminosity weighted age from a single component model
(\mage$_{\rm K10}=6.4^{+0.6}_{-0.8}$ Gyr, \citealt{Kuntschner+10}),
although different authors claim ages larger than 10 Gyr
\citep{Afanasiev+02}. The same is not true for the mean metallicity
(\mfe$=-0.04 \pm 0.03$ dex) and mean $\alpha$/Fe abundance ratio
(\malpha$=0.24 \pm 0.03$ dex) within 1$R_e$, where luminosity weighted
values found by \citet{Kuntschner+10} are more metal poor ([Z/H]$_{\rm
  K10} = -0.25 \pm 0.04$ dex) and slightly less $\alpha$ enhanced
([$\alpha$/Fe]$_{\rm K10}=0.15 \pm 0.05$ dex). The difference is
probably due to the different iron lines used to probe the total
metallicity and abundance ratio, and to the limited spectral
resolution and wavelength range of the previous work.

The two-dimensional maps of the single stellar population parameters
show a small systematic difference in the total luminosity-weighted
values of metallicity (see Tab. \ref{tab:mean_values}), but not a
significant difference in age. This is due to the fact that the
spectral \hb\ signatures are very similar for stellar ages greater
than 4 Gyr, and it is not easy to separate them with our accuracy
($\sim$ 0.2 \AA).  As in the case of NGC~3593, the stellar population
values of the two stellar disks overlap within the errorbars in some
radial bins, but systematic differences are present. This is shown by
the radial profiles in Fig. \ref{fig:rad_ssp}: the main component is more
metal rich and less $\alpha$-enhanced almost at all radii.

The main difference in age is observed in the central $5''$ (400 pc),
where \mage$_{\rm 1}(R<5'') = 7.7 \pm 0.8$ Gyr, and \mage$_{\rm
  2}(R<5'') = 5.3 \pm 0.7$ Gyr.

The main component has nearly solar metallicity at all radii
(\mfe$_{\rm 1} = -0.01 \pm 0.03$ dex), while the secondary component
varies from nearly solar at the center (\mfe$_{\rm 2}(R<5'') = - 0.03
\pm 0.03$ dex) down to subsolar outside 1.2 kpc (\mfe$_{\rm 2}(R>15'')
= -0.24 \pm 0.05$ dex). The metallicity gradient of the secondary
  component is steeper than that of the main component.

The derived mean stellar formation time scale of the main component is
$\Delta t \simeq 1$ Gyr, and it is faster for the secondary component
($\Delta t \simeq 0.3$ Gyr).

\begin{figure}
\begin{center}
\hbox{
\psfig{file=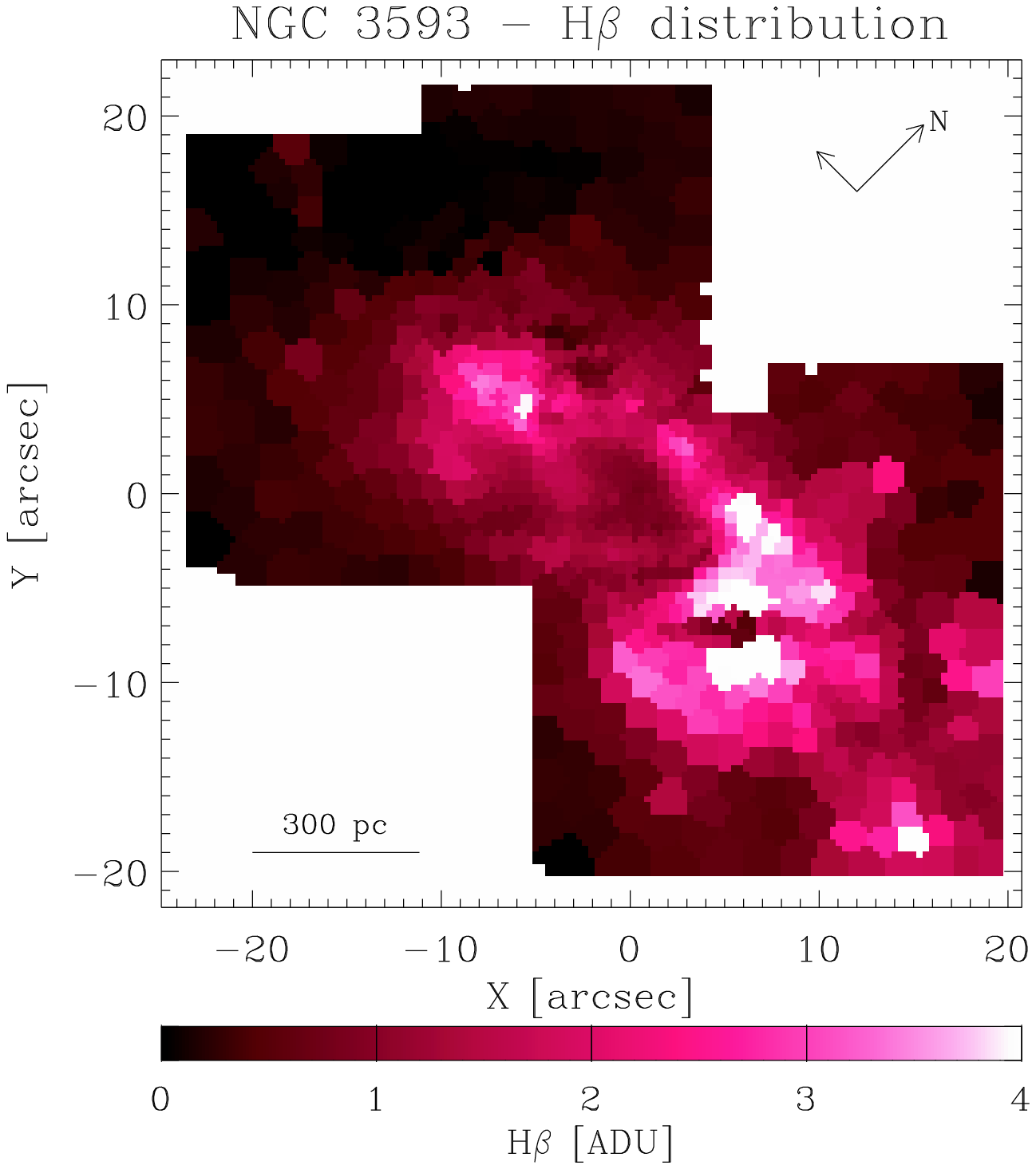,width=4.5cm,clip=}
\psfig{file=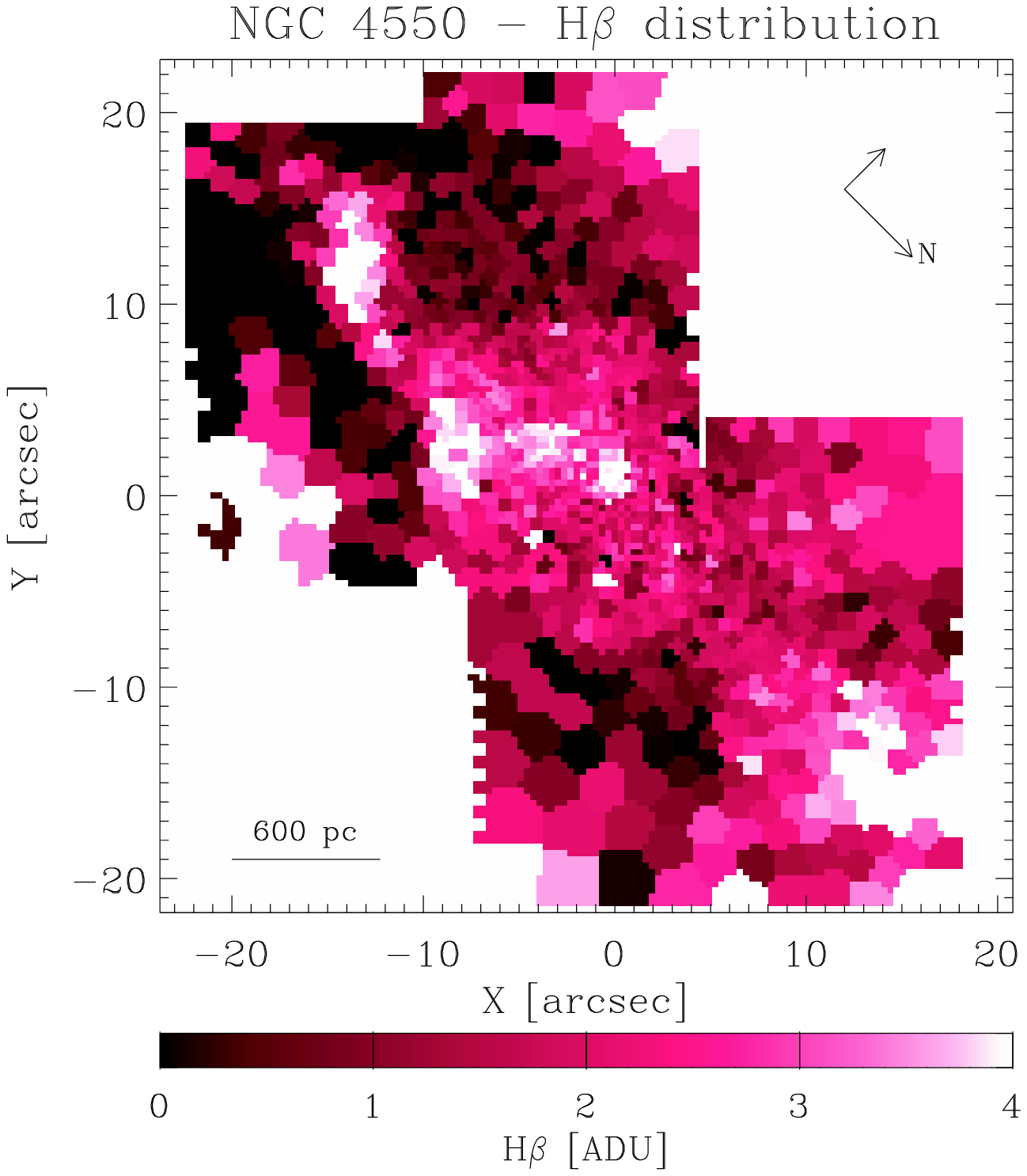,width=4.5cm,clip=}}
\caption{Intensity map of the \hb\ emission line in NGC~3593 (left
  panel) and NGC~4550 (right panel). Fluxes are in arbitrary units. An
  asymmetric ring-like structure in NGC~3593 and a central asymmetric
  concentration in NGC 4550 are clearly visible. Scale and orientation
  are given.}
\label{fig:hb_map}
\end{center}
\end{figure}

\begin{figure}
\begin{center}
\hbox{
 \psfig{file=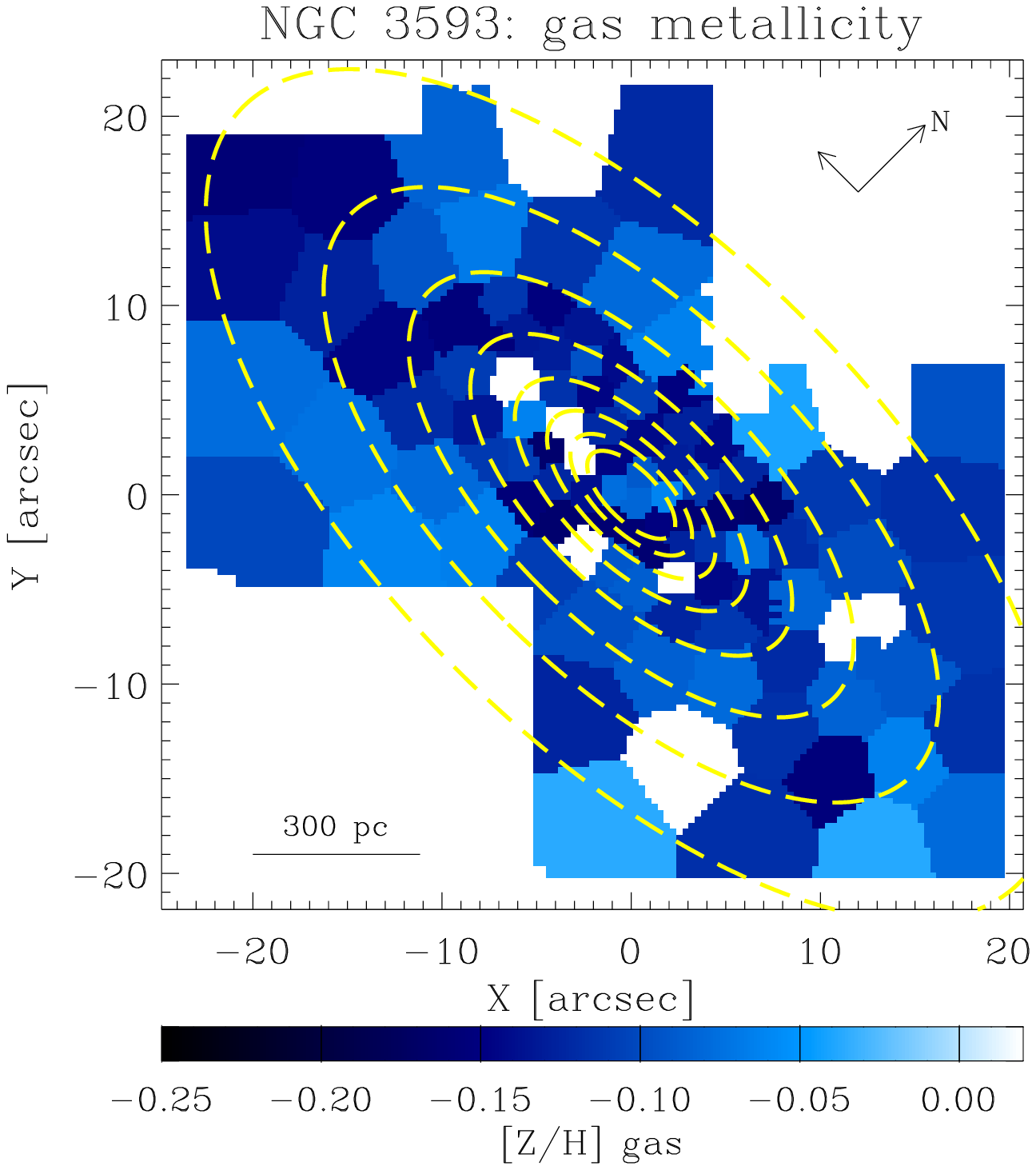,width=4.5cm,clip=}
 \psfig{file=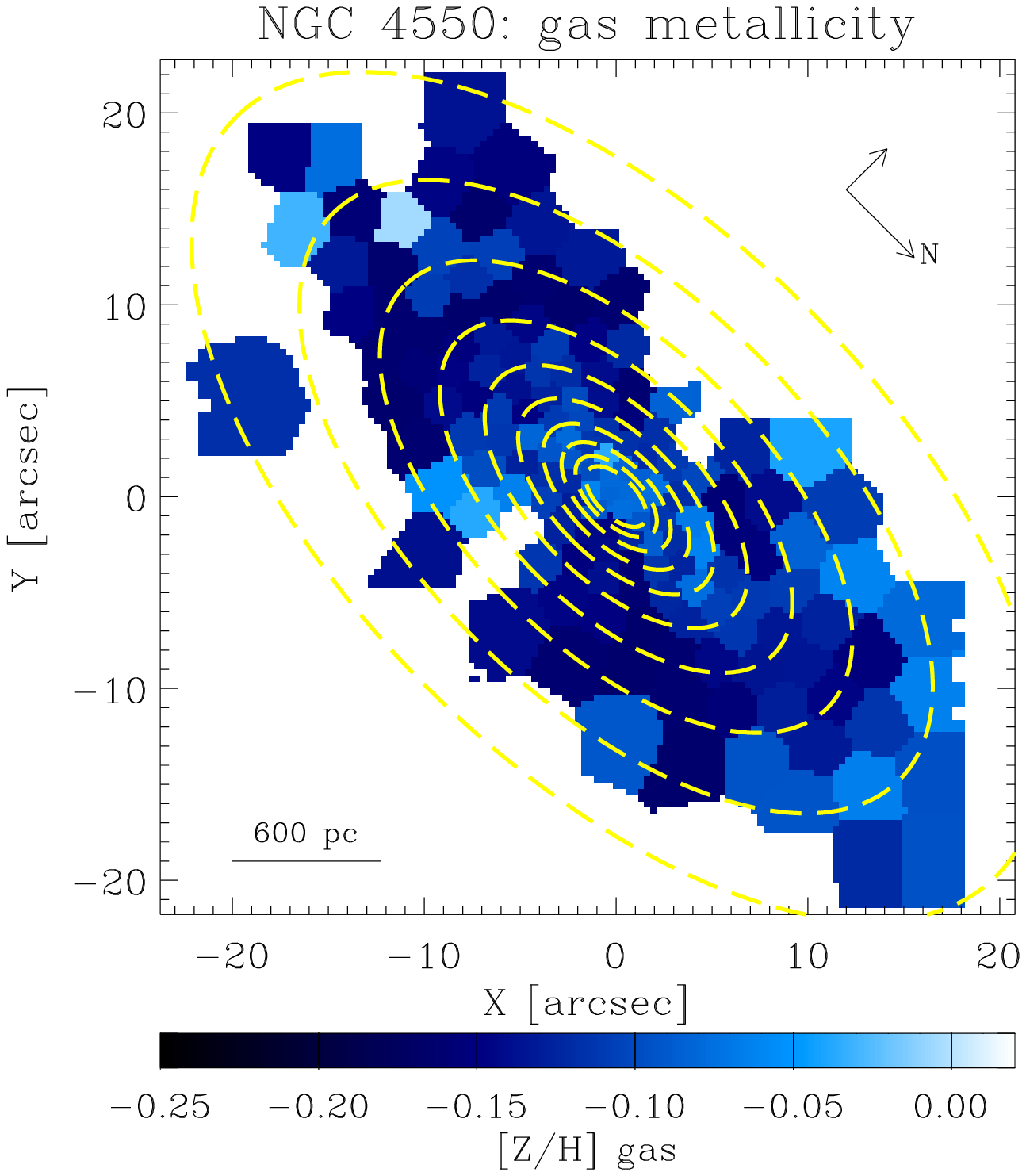,width=4.5cm,clip=}}
\caption{Map of the ionized-gas metallicity of NGC 3593 (left panel)
  and NGC 4550 (right panel). Dashed ellipses show the elliptical
  annuli used to extract the radial profiles of [Z/H]$_{\rm gas}$ shown in
  Fig. \ref{fig:rad_ssp}. Scale and orientation are given. }
\label{fig:zgas}
\end{center}
\end{figure}

\subsection{Mass surface density}
\label{sec:mass}

From the surface brightness and  stellar population properties we
can derive the mass surface density associated to each component. We
used the stellar population models by \citet{Maraston+05} to recover
the stellar mass-to-light ratios from our stellar population
measurements, and we then multiplied them for the light contribution
measured from the spectra to get the mass.  Since no flux calibration
was performed, the mass profiles are relative to the central
value. The radial profiles of stellar mass and mass-to-light ratios
are shown in Figure \ref{fig:mass_profile}.

In both galaxies, the stellar component that rotates in the same
direction of the gas is the less massive one. In NGC 3593 this is true at
all radii. In NGC 4550 this difference is evident only in the inner
$7''$ (550 pc), whereas the mass distributions of the two stellar disks are very
similar outside.

\subsection{Ionized-gas distribution}
\label{sec:ionized_gas}

The two-dimensional distributions of the ionized gas in NGC 3593 and
NGC 4550 derived from the fit to the \hb\ emission line are shown in
Figure \ref{fig:hb_map}. These maps were produced using a specific
Voronoi binning scheme with high spatial resolution, optimized to
highlight the morphology of the \hb\ distribution.

In NGC 3593, the map of the \hb\ emission displays an asymmetric
ring-like structure with a peak-to-peak diameter of $\sim 19''$ ($\sim 650$
pc), which is more intense on the W side of the galaxy. The dimension
and asymmetry of the \hb\ ring are very similar to those of the
H$\alpha$+\nii\ ring found by \citet{Corsini+98}, and to those of the
C0 ring found by \citet{Garcia+00}.

In NGC 4550, the \hb\ distribution is patchy; in particular, there is
a large double-peaked asymmetric structure in the central regions
(either a ring or a disk) with a peak-to-peak diameter of $\sim 10"$
($\sim 800$ pc). The spatial extension of this \hb\ structure is similar to
that of the regions where the mass density profile of the main disk is
higher (see Section \ref{sec:mass}) and the secondary disk is
younger (see Section \ref{sec:ssp}, right panel). The dimension of this
inner \hb\ structure is consistent with the region where CO
emission is detected \citep{Crocker+09}.

\subsection{Ionized-gas metallicity}
\label{sec:ionized_gas_metal}

The gas-phase metallicity is one of the most important observational
diagnostics of the current evolutionary state of galaxies and, in the
particular cases of NGC~3593 and NGC~4550, its measurements can give
important information on the relationship between 
the ionized-gas and the two counter-rotating stellar components.

We used the empirical relations between the intensity of the nebular
lines and metallicity to estimate the gas-phase metallicity.
Since only the \hb\/ and \oiii\/ lines are present in the spectral range we
adopted the {\it R}$_{3}$ emission-lines ratio parameter
\citep{Liang+06} defined as:

\begin{equation}
{{\it
R}_{3}}=\frac{[\ion{O}{iii}]\lambda4959+[\ion{O}{iii}]\lambda5007}{\rm
H{\small{\beta}}}.
\label{eq:r23}
\end{equation}

The galaxy spectra were not flux-calibrated. For this reason we
followed the approach suggested by \citet{Kobulnicky+03} and recently
adopted in \citet{Morelli+12} to replace the fluxes of the
emission lines with their equivalent widths. This method does not need
flux-calibrated spectra and has the further advantage of being
insensitive to reddening.
Thus, we measured the equivalent widths of the
[\ion{O}{iii}]$\lambda\lambda4959, 5007$, and \hb\/ emission lines in
the best-fit spectrum. For each emission line, we considered a central
bandpass covering the feature of interest and two adjacent bandpasses,
at the red and blue side, tracing the local continuum. The continuum
level underlying the emission line was estimated by interpolating a
straight line in the continuum bandpasses of the spectra of the
secondary component associated with the gas. The used bandpasses are
defined following \citet[Sect. 4.3]{Gonzalez+93} for \hb\/, \oiiip\/,
as done in \citet{Morelli+12}.

The gas-phase metallicity measured for NGC~3593 and NGC~4550 was
converted to [Z/H] following \citet{Sommariva+12}.  The gas
metallicity maps are shown in Figure \ref{fig:zgas}.  The
luminosity-weighted mean values along concentric ellipses were
obtained as in Sect. \ref{sec:ssp} and are plotted in
Fig. \ref{fig:rad_ssp}.

\bigskip
\noindent{\bf \it NGC 3593} 

The metallicity distribution of the gas shows an hint of a ring-like
structure of similar dimension of the \hb\ ring detected in
Fig. \ref{fig:hb_map}. The metallicities of gas and stars are
consistent for $R> 150$ pc. On the contrary, in the nucleus of the
galaxy, the gas metallicity remains constant at [Z/H]$_{\rm gas} =
-0.15 \pm 0.09$ dex, while the metallicity of the stars rises above
0.1 dex (see Sect. \ref{sec:ssp} and Fig. \ref{fig:rad_ssp}). This
discrepancy is probably due to the fact that in the very center of the
galaxy the \hb\ emission is almost absent (Fig. \ref{fig:hb_map});
therefore in the center the star formation is less efficient today,
and the present generation of stars in this region formed from a
different enriched primordial gas.

\bigskip
\noindent{\bf \it NGC 4550}

The gas metallicity distribution in NGC 4550 does not present particular
structures in its two-dimensional map. Out to $20''$ (1.5 kpc) the mean
metallicity of the ionized gas is [Z/H]$_{\rm gas} = −0.10 \pm 0.07$
dex, further out it peaks to $0.2 \pm 0.2$ dex.
The gas-metallicity radial profile is in good agreement with the
metallicity of the stellar components up to the very
center, confirming that the star-formation occurring in the gas disk is
forming the secondary component.
Contrary to NGC 3593, there are no differences between gas and stellar
metallicity in the center of the galaxy. This is due to the fact that
in this galaxy the star-formation, as traced by the \hb\ emission, is
occurring in a patchy structure all over the disk extension
(Fig. \ref{fig:hb_map}) and not mostly in the central structure.

\section{Discussion}

\label{sec:discussion}

The morphology, kinematics, and properties of the stellar populations
of NGC~3593 and NGC~4550 present both similarities and differences,
which help us in investigating their origin and assessing the most
efficient formation mechanism.

\subsection{Origin of the large-scale counter-rotating stellar disks}

In each galaxy, the main stellar disk (i.e. the most massive) hosts a
secondary stellar disk and a ionized-gas disk that counter-rotate with
respect to it. In both galaxies, the stars in the secondary disk are
on average younger, more metal poor, and more $\alpha$-enhanced than
the stars in the main disk. The differences in mass and age between
the two counter-rotating stellar disks are more pronounced in NGC~3593
than in NGC~4550. In NGC~3593 the two stellar disks have the same
intrinsic flattening but different scale lenghts, whereas in NGC~4550
the secondary component is slightly thicker than the main component,
and they have the same scale length.  Both galaxies host an asymmetric
\hb\ structure in their central kpc of similar size: it is 650 pc wide
with a ring-like shape in NGC 3593, and 800 pc wide in NGC 4550. These
\hb\ structures match the regions where the secondary stellar disks
are younger.

Our findings support the scenario of a gas accretion followed by star
formation as the origin for the stellar and gas counter-rotation in
both galaxies.

A pre-existing galaxy disk (that corresponds to the ``main''
component, as defined in Sect. \ref{sec:results}) acquired gas from
outside onto retrograde orbits. The amount of acquired gas must have
been large enough to remove the pre-existing gas. The new gas first
accreted in the outer regions of the galaxy, where formed rapidly new
stars. The star-formation time-scale in the outer regions of the
secondary stellar disk is indeed shorter than 500 Myr in both
galaxies, as inferred from the $\alpha$/Fe overabundance. Then, the
star formation continued outside-in with a longer time scale, as
inferred from the radial gradients in age and metallicity, which are
more pronounced in the secondary component than in the main
component. This process generated the secondary counter-rotating
stellar components in both galaxies. These components are indeed
younger than the corresponding main components and their angular
momenta are aligned with that of the ionized-gas disks.

The presence of molecular clouds, traced by the CO emission, which
co-rotate with both the secondary stellar disk and ionized-gas disk,
is suggestive of star formation in the secondary component (see
\citealt{Wiklind+92, Garcia+00} for NGC~3593 and \citealt{Crocker+09}
for NGC~4550). The newborn stars have on average lower metallicity
and, as a consequence of the anti-correlation between [Z/H] and
[$\alpha$/Fe], they are more $\alpha$-enhanced than the stars in the
main stellar disk. This depends on the nature of the gas accreted by
the two galaxies.

The difference between the luminosity-weighted ages of the two
counter-rotating stellar components dates the gas accretion event. In
NGC 3593 the accretion occurred between 2.0 and 3.6 Gyr ago, i.e. $1.6
\pm 0.8$ Gyr after the formation of the main stellar disk. In NGC
4550, our measurements do now allow a precise dating: it occurred
$\simeq$ 7 Gyr ago, less than 1 Gyr after the formation of the
main stellar disk.

Also the metal content of the ionized gas is consistent with the
proposed gas-accretion scenario, although the fact that the
metallicity of the gas component is similar to metallicity of {\it
  both} the stellar disks is intriguing. This may suggest that the
acquired gas mixed with the pre-existing gas, responsible for the
formation of the main component. The final mixture of gas therefore
depends on the relative enrichment and amount of the acquired and
pre-existing gas, and on the additional enrichment subsequent to the
formation of the secondary stellar component.

Although the metallicities of the ionized gas and the two stellar
disks are similar on average, in the central 250 pc of NGC~3593 we
measured a lower gas metallicity with respect to the stars
(Sect. \ref{sec:ionized_gas_metal}). This difference between the two
galaxies can be ascribed either to a different formation scenario or
to a different evolutionary stage in the formation process. In this
latter hypothesis, the gas metallicity of NGC~4550 represents the
final stage of the assembly process of the counter-rotating
disk, which is still at work in NGC~3593: the stellar disk is forming
outside-in, and the star formation will reach the center and then
disappear, leaving stellar and ionized-gas components with
approximately the same metallicity trend, as observed in
NGC~4550. This interpretation is consistent with the different ages of
the secondary stellar disks of the two galaxies.

To date, the stellar populations of counter-rotating stellar disks of
comparable sizes were successfully disentangled only in three disk
galaxies. They are: NGC~5179 \citep{Coccato+11}, NGC~3593 and NGC~4550
in this work. In all of them, the stellar component co-rotating with
the gas is younger, less massive and has different composition than
the main stellar component, in agreement with the gas-accretion
scenario.
In NGC 5719, the gas-accretion event is traced by a stream of neutral
hydrogen that connects the galaxy with its neighbor NGC~5713
\citep{Vergani+07}. Although we do not have the same evidence for NGC
3593, its stellar populations are very similar to those of NGC 5719;
this makes us confident that these two galaxies share the same
formation mechanism. The situation is not so clear for NGC 4550, as
the differences in the stellar population properties between the two
stellar counter-rotating disks are not as evident as in NGC~3593.

The scenario of internal origin of the secondary counter-rotating
components is ruled out, as it would predict the same mass and stellar
population for both components. The internal-origin scenario is ruled
out even if we include a rejuvenation process for the secondary
component (i.e. in the inner 850 pc of NGC~3593, and in the inner 400
pc of NGC~4550). In this alternative case in fact, the secondary
stellar component is expected to have higher metallicity (as the
result of star formation from enriched gas) and lower
$\alpha$-enhancement (as the result of the increased star-formation
time scale, due to the coexistence of old and young stars) than the
main stellar component in the regions where younger ages are measured.
All these predictions are not consistent with our observations, and
therefore we discard such a scenario.

On the contrary, a binary galaxy merger cannot be completely ruled out
by our measurements as the formation mechanism of counter-rotating
stellar disks, given the limited galaxy sample and given that we have
conclusive evidence of gas accretion only for one case. According to
\citet{Crocker+09}, the merging scenario explains the fact that the
gas rotates together with the thicker component in NGC 4550. On the
other hand, we show that the difference in thickness between the two
counter-rotating stellar disks in NGC 4550 is not as pronounced as the
one that can be guessed without a proper spectral decomposition. In
addition, we found that the thicker stellar component is the youngest,
and therefore it is the accreted component, whereas in the
simulations the thicker component is the pre-existing stellar disk
that was dynamically heated by the merging process \citep{Crocker+09}.

\subsection{The most efficient mechanism to form large-scale counter-rotating stellar disks}

We know that {\it both} binary mergers and episodes of gas accretion
do occur in our Universe, therefore they are both viable mechanisms to
create large-scale counter-rotating stellar disks. However, the
relative fraction of counter-rotating components built by these
processes is still unknown.

As stated in the introduction, in the gas-accretion scenario the
stellar component associated with the ionized gas is predicted to be
always younger than the main stellar component. In binary galaxy
mergers instead, one would expect that the younger stellar component
is associated with the ionized gas only in 50 per cent of the
cases. Let's define $M$ as the fraction of galaxies hosting two
counter-rotating stellar disks of comparable sizes that were generated
by a binary major merger (thus, $1-M$ is the fraction produced by gas
accretion), and $E$ as the event of finding the stellar component
co-rotating with the gas to be the youngest.
 
The probability $P_E$ to observe $E$ is:

\begin{equation}
  P_E(M)=1-M+M/2 = 1-M/2
\end{equation}

The probability $\Pi_E$ to observe $E$ in exactly $N$ galaxies out of a
sample of $T$ counter-rotating galaxies is the first term of the Binomial
distribution:

\begin{equation}
\Pi_E(N,T,P_E(M)) = {T \choose N} P_E^N \left(1-P_E \right)^{T-N}
\end{equation}

The value of $M$ for which $\Pi_E$ is maximum gives us the most probable
value of the fraction of counter-rotating galaxies produced by merger
events. The
condition:

\begin{equation}
\int^{M+M^+}_{M-M^-} \Pi_E(N,T,P_E(x)) dx = 0.67 \cdot \int^{1}_{0} \Pi_E(N,T,P_E(x)) dx
\end{equation}
defines the $1\sigma$ error bar on $M$. In our case ($T=N=3$) we find
that the fraction of counter-rotating galaxies built by binary mergers
is $M < 44$ per cent. Therefore, measuring the difference in age of
the counter-rotating components in a larger sample of counter-rotating
galaxies is fundamental to identify the most efficient
mechanism. Large spectroscopic surveys like Atlas3D
\citep{Cappellari+11} or CALIFA \citep{Sanchez+12} will help to
identify other potential candidates by recognizing the kinematic
signatures of stellar counter-rotating disks, like two symmetric peaks
in the velocity dispersion map \citep{Krajnovic+11}.

\section{Summary}
\label{sec:summary}

We presented the VIMOS/VLT two-dimensional kinematics of the stars and
the ionized-gas in the two disk galaxies NGC~3593 and NGC~4550, which
are known to host a stellar and a ionized gas disks that are
counter-rotating with respect the main galaxy disk.

We applied the novel spectroscopic decomposition technique introduced
in \citet{Coccato+11} to separate and measure from
  two-dimensional spectroscopic data the morphologies, the
kinematics, the stellar populations, and the mass surface densities of
{\it both} the two counter-rotating stellar disks in the sample
galaxies for the first time.

We found that in both galaxies the two stellar disks have different
stellar population properties, as measured by the equivalent width of
the \hb, Mg and Fe absorption lines. The difference is more evident in
NGC 3593 than in NGC 4550.  Both galaxies host a counter-rotating
stellar disk, which rotates along the same direction of the ionized
gas, and which is on average less massive, younger, metal poorer,
  and more $\alpha$ enhanced than the main stellar
galaxy disk. Our results support the scenario of external gas
acquisition, followed by a subsequent outside-in star formation as the
origin of the observed counter-rotation. In NGC 3593, we are able to
date the acquisition event that formed the counter-rotating stellar
disk ($1.6 \pm 0.8$ Gyr after the formation of the main galaxy disk),
while for NGC 4550 we can set only an upper limit (less than 1 Gyr
after the formation of the main galaxy disk).

Our results, combined with previous studies in the literature, rule
out the internal scenario as the origin of counter-rotation in the studied
galaxies. On the contrary, the merger scenario cannot be completely
ruled out, given the small statistic available. Thus, a larger sample
is necessary to identify the most efficient mechanism.

\begin{acknowledgements} 
  We wish to thank the referee G. Worthey for his useful comments that
  helped to improve this paper.  LC acknowledges financial support
  from the European Community’s Seventh Framework Programme
  (/FP7/2007-2013/) under grant agreement No. 229517. The work has
  been partially founded by the grants 60A02-1283/10, 60A02-5052/11,
  60A02-4807/12 of the Padua University.

\end{acknowledgements}

\bibliography{coccato12}
                                           
\end{document}